%% file: main.tex
\documentclass[10pt,letterpaper]{article}
\usepackage[top=0.85in,left=2.75in,footskip=0.75in]{geometry}
\pdfoutput=1 
\usepackage{amsmath,amssymb}
\usepackage{subcaption}

\usepackage{changepage}

\usepackage[utf8x]{inputenc}

\usepackage{textcomp,marvosym}

\usepackage{cite}
\usepackage{csquotes}

\usepackage{nameref,hyperref}

\usepackage[right]{lineno}

\usepackage{microtype}
\DisableLigatures[f]{encoding = *, family = * }

\usepackage[table]{xcolor}

\usepackage{array}

\newcolumntype{+}{!{\vrule width 2pt}}

\newlength\savedwidth

\newcommand\thickhline{\noalign{\global\savedwidth\arrayrulewidth\global\arrayrulewidth 2pt}%
\hline
\noalign{\global\arrayrulewidth\savedwidth}}

\raggedright
\usepackage[aboveskip=1pt,labelfont=bf,labelsep=period,justification=raggedright,singlelinecheck=off]{caption}

\makeatletter
\renewcommand{\@biblabel}[1]{\quad#1.}
\makeatother

\usepackage{lastpage,fancyhdr,graphicx}
\usepackage{epstopdf}
\usepackage{booktabs}
\fancyheadoffset[L]{2.25in}
\fancyfootoffset[L]{2.25in}
\begin{document}
\vspace*{0.2in}

\begin{flushleft}
{\Large
\textbf\newline{Misinformation, Believability, and Vaccine Acceptance Over 40 Countries: Takeaways From the Initial Phase of The COVID-19 Infodemic} 
}
\newline
\\
Karandeep Singh\textsuperscript{1\Yinyang},
Gabriel Lima\textsuperscript{2,1\Yinyang},
Meeyoung Cha\textsuperscript{1,2*},
Chiyoung Cha\textsuperscript{3},
Juhi Kulshrestha\textsuperscript{4},
Yong-Yeol Ahn\textsuperscript{5,6,7},
Onur Varol\textsuperscript{8}
\\
\bigskip
\textbf{1} Institute for Basic Science, Daejeon, 34126, South Korea
\\
\textbf{2} School of Computing, KAIST, Daejeon, 34141, South Korea
\\
\textbf{3} Ewha Womans University, Seoul, 03760, South Korea
\\
\textbf{4} GESIS - Leibniz Institute for the Social Sciences, Unter Sachsenhausen 6-8, 50667 Cologne, Germany
\\
\textbf{5} Center for Complex Networks and Systems Research, Luddy School of Informatics, Computing, and Engineering, Indiana University, IN 47408, USA
\\
\textbf{6} Indiana University Network Science Institute, Indiana University, IN 47408, USA
\\
\textbf{7} Connection Science, Massachusetts Institute of Technology, MA 02139, USA
\\
\textbf{8} Faculty of Engineering and Natural Sciences, Sabanci University, Istanbul 34956, Turkey
\\
\bigskip

\Yinyang These authors contributed equally to this work.

* mcha@ibs.re.kr

\end{flushleft}

\newcommand{\mc}[1]{\textcolor{magenta}{#1}}
\newcommand{\cc}[1]{\textcolor{brown}{#1}}
\newcommand{\gl}[1]{\textcolor{blue}{#1}}
\newcommand{\ks}[1]{\textcolor{purple}{#1}}
\newcommand{\ov}[1]{\textcolor{blue}{#1}}
\newcommand{\jk}[1]{\textcolor{cyan}{#1}}
\newcommand{\deindent}[1]{\hspace{-1.0em}#1}

\section*{Abstract}
\input{content/0abstract}


\section*{Introduction}
\input{content/1intro}

\section*{Materials and Methods}
\input{content/2methods}

\section*{Results}
\input{content/3_0results}

\input{content/3_1vacc}

\section*{Discussion}
\input{content/4discussion}

\section*{Conclusion}
\input{content/5conclusion}


\newpage
\bibliography{refs}

\newpage
\section*{Supporting Information}
\input{content/6si}
\input{content/model_results}

\end{document}

%% file: content/0abstract.tex
The COVID-19 pandemic has been damaging to the lives of people all around the world. Accompanied by the pandemic is an \textit{infodemic}, an abundant and uncontrolled spreading of potentially harmful misinformation. The infodemic may severely change the pandemic's course by interfering with public health interventions such as wearing masks, social distancing, and vaccination. 
In particular, the impact of the infodemic on vaccination is critical because it holds the key to reverting to pre-pandemic normalcy. This paper presents findings from a global survey on the extent of worldwide exposure to the COVID-19 infodemic, assesses different populations' susceptibility to false claims, and analyzes its association with vaccine acceptance. 
Based on responses gathered from over 18,400 individuals from 40 countries, we find a strong association between perceived believability of misinformation and vaccination hesitancy. Additionally, our study shows that only half of the online users exposed to rumors might have seen the fact-checked information. Moreover, depending on the country, between 6\% and 37\% of individuals considered these rumors believable. Our survey also shows that poorer regions are more susceptible to encountering and believing COVID-19 misinformation. 
We discuss implications of our findings on public campaigns that proactively spread accurate information to countries that are more susceptible to the infodemic. We also highlight fact-checking platforms' role in better identifying and prioritizing claims that are perceived to be believable and have wide exposure. 
Our findings give insights into better handling of risk communication during the initial phase of a future pandemic.

%% file: content/1intro.tex
In the contemporary world with social media, misinformation and disinformation can be rapidly disseminated to millions of people~\cite{kwon2017plos,shao2018spread}. Studies suggest that harmful false information spreads more broadly than the truth online~\cite{science18vosoughi}. Due to social media's global reach with a rapid amplification mechanism~\cite{flickr2009www}, information can quickly inundate the Internet and get reinforced, potentially creating an ``infodemic''~\cite{zarocostas2020fight,who20infodemic}. This abundance of information can lead to harmful consequences. For instance, the COVID-19 infodemic has resulted in seemingly harmless acts such as eating vegetables, shaving one's head, or saltwater gargling~\cite{kh20saltwater} to norm-violating and damaging acts like arson~\cite{nyt20burning}.

Previous work addressing misinformation in healthcare has found that false and misleading claims negatively influence people's attitudes towards vaccine acceptance. One study conducted in the Democratic Republic of the Congo during the Ebola epidemic found an adverse effect of false information on vaccine acceptance~\cite{vinck2019institutional}. The World Health Organization (WHO) has highlighted how misinformation has raised doubts on the effectiveness of human papillomavirus (HPV) vaccines~\cite{whohpv}. Furthermore, vaccine refusal has led to the measles' resurgence in the US, even after decades of containment~\cite{benecke2019anti}. Research has proven that vaccines do save lives~\cite{orenstein2017simply}. In the context of the coronavirus pandemic, a vaccine is widely believed to be the only way out towards pre-pandemic normalcy~\cite{draulans2020finally}. 


Due to the Internet's nature, it is challenging to prevent the spread of false information~\cite{kwon2013icdm}. There is no established authority that checks the veracity of the information that is shared. Moreover, social media can quickly spread a piece of information to large groups of people, independently of its source and authenticity. Misinformation, disinformation, and eccentric opinions can get reinforced by repeated exposure and threaten public health. As a result, communicating even the most basic facts to the public can become a challenge in itself. 

\begin{figure}[htpb!]
\caption{\textbf{Distribution of study participants around the world. The study obtained responses from 18,407 participants from 40 countries.}}
\vspace{2mm}
    \centering
    \includegraphics[width=.9\linewidth]{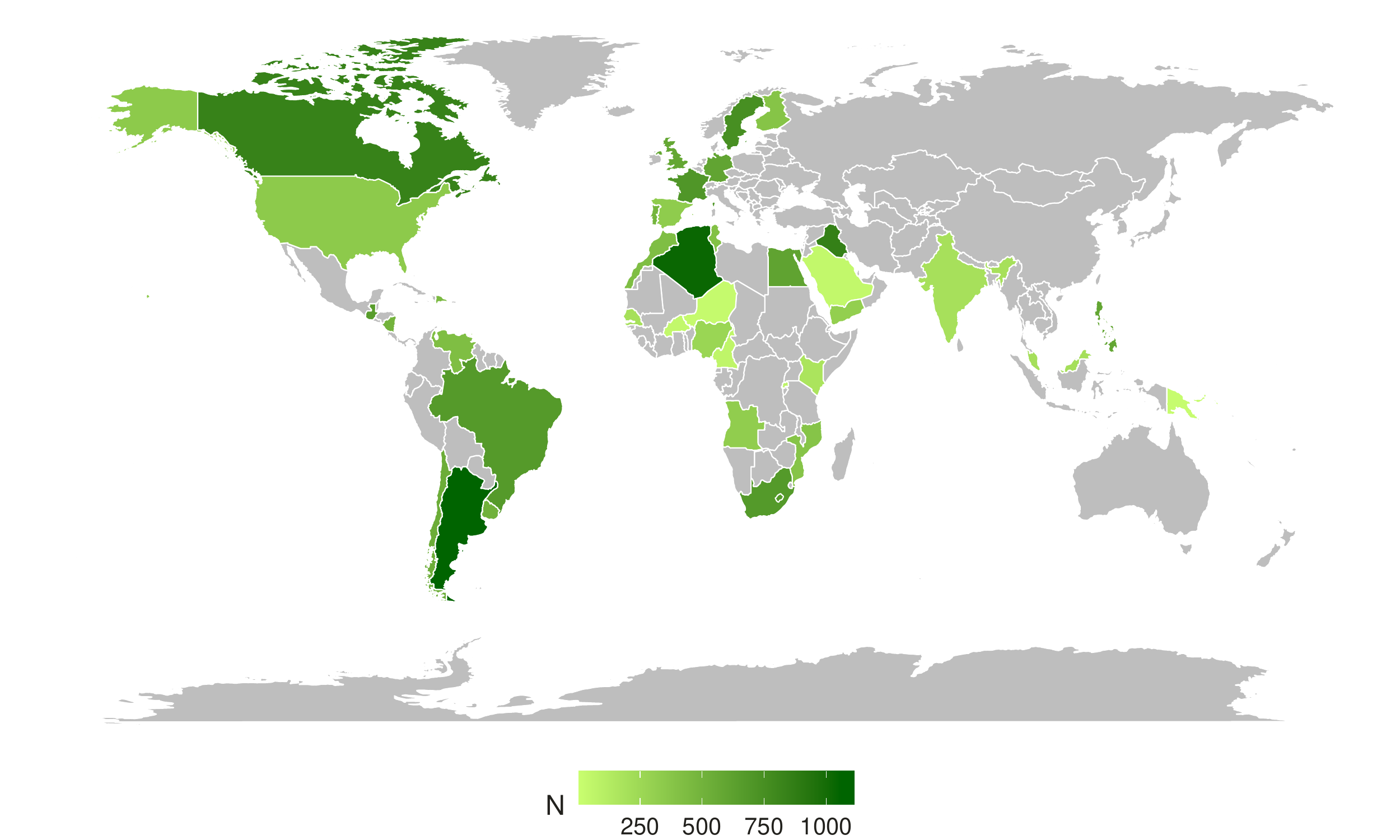}
    
    \label{fig:map}
\end{figure}

A possible remedy to the harm caused by an infodemic is flagging and removing false information from the Internet and social media. Extensive research has focused on automating this process~\cite{ma2016ijcai,ma2018rumor}, and recently, social media platforms have taken both proactive and reactive steps to prevent and minimize the spread of misinformation~\cite{nyt20twitter,verge20qanon}. 

Proactive dissemination of fact-checked information preempting the spread of the infodemic is another way to combat misinformation. Working on these lines, we had launched an online campaign to debunk COVID-19 rumors~\cite{fbr} that disseminated accurate coronavirus-related information reaching over 50,000 individuals. The campaign aimed to collect fact-checked information from regions that had already suffered from the infodemic and spread them to other regions where the infodemic was at its infancy.

Many studies have addressed the pressing issue of of COVID-19 infodemic~\cite{info20global, twit20explora, cinelli2020covid, kouzy2020coronavirus}. These studies target the problem either from a computational or exploratory perspective. In this work, we deploy a descriptive survey based approach to analyze the prevalence of infodemic. We utilize the data gathered from our public campaign and study the public susceptibility to the infodemic by analyzing the individual survey responses. Through a global-scale survey conducted using the Facebook Advertisement Platform, we utilize 18,407 complete responses from individuals in 40 countries (see Fig~\ref{fig:map}) and measured the extent to which a wide range of coronavirus-related rumors and their respective fact-checks reached different countries. We also examined the impact of exposure to misinformation on attitudes towards rumor believability and vaccine acceptance.

We find that distinct claims disseminate differently around the world, disproportionately affecting less economically developed countries. Our results also indicate that exposure to false information is nearly two times more prevalent than that of corresponding fact-checked information, indicating that half of the online users exposed to rumors might not have encountered the corresponding fact-checks. 

When jointly considering believability, the picture becomes more complicated. Although some claims---such as those 5G-related ones---exhibit inherently low believability, we find that they spread widely. Other popular claims, e.g., regarding the use of existing drugs, have relatively high believability. This finding implies that fact-checking organizations could utilize user response or quick polls to identify claims that are more likely to be widely believed. Such a prioritization strategy could be helpful given the limited resources at fact-checking systems.

Our study reflects a positive association between exposure and both believability and vaccination hesitancy. The results also show that those who perceive the pandemic as more threatening are more willing to accept a future vaccine, highlighting the importance of raising public awareness concerning the disease's risks. Our regression analysis suggests that exposure to fact-checks could nearly balance out the adverse effect of exposure to misinformation. This remedy, however, doesn't seem to be effective for individuals who report being susceptible to false information; to what extent participants found claims believable was much more strongly associated with vaccine hesitancy. 

Given the rising social media usage, including in the developing and underdeveloped regions, social media platforms could be used as a primary medium for disseminating fact-checks. We propose one algorithmic prioritization method for future debunking strategies that account for the varying degrees of believability and exposure of claims. This work demonstrates how web data can be analyzed to understand important health implications during a global pandemic. We describe our methodology for conducting surveys over a social media platform and post-processing the data to correct for sampling demographic biases.  

%% file: content/2methods.tex

\subsection*{COVID-19 Claims Selection}
For identifying popular false claims, we collected over 200 COVID-19 rumors from DXY.cn, a Chinese online community for physicians and healthcare professionals, on March 18, 2020. This site hosted a comprehensive list of Chinese social media rumors during the COVID-19 infancy in China. Many of them were later found to have spread worldwide. After removing redundant content and lockdown-related claims, we investigated 30 pieces of misinformation addressing health-related behaviors. We combined these pieces into 11 distinct claims (e.g., combining the effect of multiple different rumors into a single claim). We also categorized claims into subgroups depending on the rumor' nature (e.g., those addressing vaccination or do it yourself (DIY) measures). Corroborating these claims with fact-checked information from credible sources, including the World Health Organization (WHO) 's Mythbusters~\cite{whomythbusters} and the International Fact-Checking Network (IFCN) 's \#CoronaVirusFacts Alliance database~\cite{covidpoynter}, we arrived at the following list of 11 misinformation claims: 

\begin{enumerate}
    \item \textsf{5G} (5G): 5G networks can contribute the spreading of the coronavirus.
    \item \textsf{Dryer} (Hot\&Co): Hot-air dryers can kill the coronavirus.
    \item \textsf{Gargling} (DIY): Gargling with salt water can prevent coronavirus infection.
    \item \textsf{Drugs} (DIY): Existing drugs for malaria and HIV can help treat COVID-19.
    \item \textsf{Pharma} (Vaccination): Pharmaceutical companies are spreading COVID-19 so they can profit from its vaccine.
    \item \textsf{Population} (Vaccination): The COVID-19 vaccines currently being developed are forms of population control.
    \item \textsf{Sunbath} (Hot\&Co): Standing in the sun can kill the coronavirus.
    \item \textsf{Tracking} (Vaccination): The COVID-19 vaccine is being developed to implant people with tracking microchips.
    \item \textsf{Vinegar}(DIY): Apple cider vinegar can kill the coronavirus in the throat.
    \item \textsf{Water} (DIY): Drinking water every 15 minutes will prevent getting infected with the coronavirus.
    \item \textsf{Weather} (Hot\&Co): The coronavirus will only spread in cold, dry weather and does not survive in hot, humid weather.
\end{enumerate}

Our survey and claims were translated into English, French, Spanish, Portuguese, and Arabic. For French and Spanish, we first used Google Translate to obtain crude translations and then used the Prolific crowdsourcing platform~\cite{palan2018prolific} to recruit native speakers from these languages (minimum 18 each) to refine the translations. Recruited participants attended a short survey in which they were asked to refine the provided translations. This procedure was repeated three times for each language in an iterative manner. The translations were done entirely by volunteering native speakers from the second author's institution for Portuguese and Arabic.

\subsection*{Survey Design}







This study had been approved by the Institutional Review Board at the corresponding author’s institution (KAIST IRB-20-229) and was performed in accordance with the relevant guidelines and regulations. Informed consent was obtained from all study participants. Participants were asked their current residence country and their level of worry regarding the pandemic using a 5-pt Likert Scale. 
Our survey was designed to address four different aspects of the coronavirus infodemic to the public: \textit{i}) exposure to misinformation, \textit{ii}) exposure to fact-checks, \textit{iii}) perception of claim believability, and \textit{iv}) perception of how beneficial fact-checks could be to one's community. The following questions were presented in random order to the survey participants for each of the claims:

\begin{enumerate}
    \item \textit{Exposure}: Have you seen or heard this information in the last month? ~~(Answer: Yes, Partly, No, I don't remember)
    \item \textit{Fact-Checks}: Have you ever seen an official source confirming or denying the claim above? ~~(Answer: Yes, No, I don't remember)
    \item \textit{Believability}: How believable does the information above seem to you? ~~(Answer: Not believable at all, Not really believable, I am not sure, Somewhat believable, Very believable)
    \item \textit{Benefit}: To what extent would your community benefit from seeing a fact-checking result of the claim above? ~~(Answer: Not at all, A little, Moderately, A lot)
\end{enumerate}

Participants were not explicitly asked whether they believed the study's rumors to avoid potential social desirability biases. Respondents might have answered that they did not believe the claims so that they would be seen more favorably. Hence, we phrased our questions such that they were asked to what extent they found the rumors to be believable. At the end of the survey, respondents were also asked demographic questions and to what extent they perceived the novel coronavirus as a threat, which was measured via a threat scale introduced in~\cite{kachanoff2020realistic} (termed \textit{perceived threat}).

\subsection*{Data Collection}

We conducted a large-scale online survey using the Facebook Advertising Platform from June 18 to July 13, 2020. The survey was designed in the SurveyMonkey platform, and the link to the survey was made available via advertisements on Facebook. As of March 2020, Facebook had 2.60 billion monthly active users and 1.73 billion daily active users~\cite{facebookreport}, making it the largest social media platform. Some recent publications~\cite{schneider2019s,pham2019online} have explored Facebook's usage as a survey platform and noted its advantages of a deep and broad reach, rapid data collection, granular targeting, and cost-effectiveness. The Facebook Advertising Platform allows targeting based on age, location, spoken language, and interests. Sample biases can be dealt with the adequate application of post-stratification weighting techniques, although studies such as~\cite{ribeiro2020biased} show that the demographic distributions of Facebook users tend to not differ hugely from census distributions.

To obtain a large and more representative sample of every country, we designed independent Facebook campaigns for the target countries. Each campaign was further divided into four advertisement sets by age groups (18-24, 25-44, 45-64, and 65+ years). The English survey was run from June 18 to June 25, the Portuguese, Spanish and French surveys were conducted from June 22 to June 28, while the Arabic survey was run from July 7 to July 13. 

To control for demographic factors, we ensure a minimum sample of 100 responses from each country. For countries with less than 30 complete responses from the initial round, the survey was rerun from July 7 to July 13. Since the respondents were recruited through the Facebook Advertising Platform, they did not receive any financial benefit from participating in the survey. Therefore, participation is voluntary, and respondents could choose to withdraw from the survey at any time. 

We obtained 1,946,516 responses ($N$=44,239) from Facebook users who have seen and clicked on our advertisement. We discarded incomplete responses and participants with duplicated IP addresses. Due to our weighted analysis, which requires each participant to report their sex, age, and country of residence, we discarded responses from participants who chose not to reveal their sex. We allowed participants to report their sex as ``other.'' To weigh responses successfully, we solely kept participants from those countries with at least 30 complete responses. Our final dataset consisted of 805,816 complete responses ($N$=18,314) from 40 countries (see~Fig.~\ref{fig:map}). Our sample covers all continents and contains a median of 464 respondents per country. Demographic information regarding all participants is presented in SI Table S1.

\subsection*{Sample and Weighting}

Recruiting participants through the Facebook Advertising Platform allowed us to reach a larger and more representative pool of respondents than otherwise possible through crowdsourcing platforms. Nevertheless, Facebook users are still not demographically representative of countries' populations. For instance, although previous work has found a high correlation between the US Census and Facebook users, the latter was composed of younger and more educated people~\cite{ribeiro2020biased}. 

To compensate for any imbalance between our sample and the general population demographic distributions, we employed \emph{raking} as a post-stratified weighting technique that assigns a weight to each response according to its respondent's demographics. Raking is an iterative method that calculates weights for each joint demographic group concerning separate demographic distributions until convergence~\cite{kalton2003weighting}. These weights can be used to estimate a population's information more accurately given a non-representative sample. Previous research has shown that Facebook user demographics are comparable to gold-standard surveys and the differences can be dealt with by appropriate weighing techniques~\cite{fb2020biased}.

After obtaining estimates of each country's age and sex distributions from the United Nation's 2019 World Population Prospects dataset, we used raking for calculating each response's weight. This technique was employed for each country, resulting in weights for each sex, age group, and country triple in our dataset. As the UN provides information about age distributions in 5-year intervals, we used the 20-24 bracket corresponding to our 18-24 age group for weighting purposes. Unless stated otherwise, we present weighted results from our analysis.

\begin{figure}[htb!]
\caption{\textbf{Sample and population age distributions for Brazil ($N$=698).}} 
As an example, this figures highlights the differences between weighted and unweighted exposure rates to the \textsf{5G} claim. Error bars are standard errors.

    \centering
    \vspace{3mm}
    \includegraphics[width=.8\linewidth]{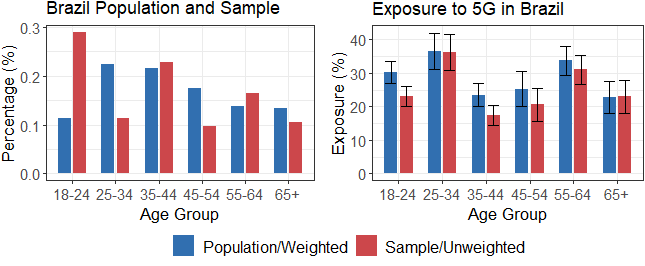}
    
    \label{fig:methods}
\end{figure}

Figure~\ref{fig:methods} exemplifies the weighing process. The left plot shows our sample's age distribution from Brazilian survey respondents from Facebook alongside the true Brazilian population's age distribution. By comparing the two, it is possible to note that the survey sample is younger than the real population. Weighting techniques like raking compensate for this discrepancy and help obtain a more accurate population-level estimation~\cite{kalton2003weighting}. The right plot in Figure~\ref{fig:methods} shows how weighting adjusted the sample's estimates for the case of exposure to the \textsf{5G} rumor. Some age groups exhibit statistically significant differences between the weighted and unweighted estimates.

\subsection*{Regression Models}
This work presents three different regression models, all of which consider reported demographics features and the respondents' mean perceived threat as control variables. Demographic features in this study are reported age group, sex, education, health, and financial status. Except for sex and vaccination history-related dummies, all variables are treated as continuous or counts. Independent variables correspond to exposure to misinformation, and the respective fact-check counts for Model 1, and additionally, average believability for models 2 and 3. We add interaction terms for vaccination history (as history for a non-mandatory vaccine implies past vaccination) as well for exposure to claims and their respective fact-checks (to be exposed to a fact-checked also implies that one has been exposed to the claim, even if only at the time of debunking). Table~\ref{tab:codes} presents how variables were coded in all models.

Model 1 is a linear regression predicting the average reported believability of false claims. It is of the form Equation 1

\begin{equation}
    Y \sim \alpha + \sum_{i}\beta_i.C_{i} + \sum_{g}\beta_{g}.I_{g} + \sum_{k}\beta_k.T_k \label{eqn:belief}
\end{equation}
%
where $Y$ is the mean believability, $\alpha$ is the intercept term, and $\beta_i$, $\beta_g$, $\beta_k$ are coefficients for control variables $C_{i}$, independent variables $I_{g}$, and interaction terms $T_k$ respectively. 

Model 2 and 3 are logistic regression models predicting the dichotomized responses to the COVID-19 vaccine acceptance question (with the third option ``I don't know'' treated as a negative response). In order to segregate and identify the association of different categories of claims on the vaccination acceptance, we divide the claims into four groups in Model 3: DIY, Hot\&Co, vaccination conspiracies, and the 5G-related claim. For Model 3, the independent variables are distinct exposure and fact-check counts and the mean believability of the segregated groups, whereas, for Model 2, we aggregate these variables across all claims. Also, we utilize respondents' vaccine history as another control variable alongside the respondent's perceived threat. Model 2 and 3 are of the form Equation 2

\begin{equation}
    \log{\frac{P}{1 - P}} \sim \alpha + \sum_{i}\beta_i.C_{i} + \sum_{c, g}\beta_{cg}.I_{cg} + \sum_{h}\beta_h.V_{h} + \sum_{k}\beta_{k}.T_{k}  \label{eqn:vacc}
\end{equation}

where $P$ is vaccine acceptance, $\alpha$ is the intercept term, $\beta_i$, $\beta_{cg}$, $\beta_h$, $\beta_{k}$ are coefficients for control variables $C_{i}$, independent variables $I_{cg}$ (representing question categories $c$ and variable $g$), vaccination history $V_h$, and interactions terms $T_{k}$ (as explained above), respectively. We also run the same models with country-level random effects, with lasso and elastic regularization and present the results in the SI Tables S3 - S14. 

%% file: content/3_0results.tex
\begin{figure*}[t]
\begin{adjustwidth}{-2.25in}{0in}
\caption{\textbf{Country-level exposure to rumors and fact-checks.
The pink polygon presents the weighted percentage of people who have been exposed to rumors. The purple polygon shows exposure to fact-checks.}}
\vspace{2mm}
    \centering
    \includegraphics[width=\linewidth]{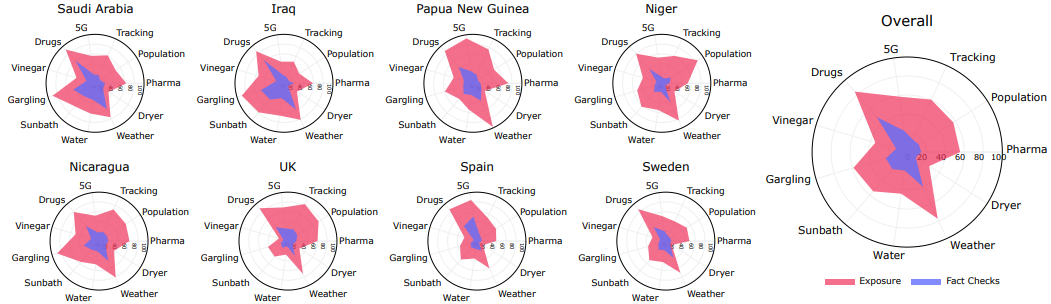}
     
    \label{fig:countries}
\end{adjustwidth}
\end{figure*}

\subsection*{Exposure to Rumors}

Figure~\ref{fig:countries} shows to what extent different countries have been exposed to COVID-19 misinformation and their respective fact-checks. The countries presented in the figure were selected to cover different regions of the world and varying exposure levels to the claims. We also show the overall weighted average across all 40 countries. All others countries are presented in SI Fig S1.

When we examine the overall exposure, we note that some claims have exceptionally high public appeal.
\textsf{Drugs} and \textsf{Weather} are the two most popular claims in our data, with an average of 84.2\% and 77.1\% of participants encountering these rumors, respectively. Vaccination-related claims show moderate-to-high exposure with \textsf{Tracking} seen by 60.7\%, \textsf{Population} by 57.6\% and \textsf{Pharma} by 55.7\% of the respondents. The smallest portion of respondents were exposed to do-it-yourself (DIY) rumors on preventive measures with \textsf{Sunbath} seen by 54.7\%, \textsf{Water} by 44.3\%, \textsf{Vinegar} by 34.9\%, and \textsf{Dryer} by 27.4\% of the respondents.
 
Next, we investigated the COVID-19 infodemic reach across different geographical regions by comparing the pink polygons in the figure representing the extent to which a country has been exposed to the infodemic rumors. Countries in the same region are exposed to a similar extent to the rumors (exemplified by a similar shape of the polygon for the Middle Eastern countries of Saudi Arabia and Iraq, or the European nations of the UK, Spain, and Sweden). At the same time, there is a noticeable variance in the infodemic's reach across continents. 

Moreover, the selected claims demonstrate varying levels of exposure in different countries. Some rumors are regionally concentrated. For instance, \textsf{Gargling} was widely disseminated in Saudi Arabia, Iraq, and Nicaragua (i.e., above 90\%), with markedly low exposure in the UK, Spain and Sweden (i.e., 40\% or less). Similarly, \textsf{5G} was seen widely in Papua New Guinea, Spain, and the UK, but less in other regions. On the other hand, rumors regarding \textsf{Drugs}, \textsf{Weather}, and vaccine-related claims (\textsf{Tracking}, \textsf{Population}, and \textsf{Pharma}) spread globally and received significant exposure across countries.

\subsection*{Exposure to Fact-Checks}

The inner purple polygons in Figure~\ref{fig:countries} show the extent to which participants were exposed to fact-checks for each claim covered by our study. Our results reveal that 48.6\% of participants were exposed to fact-checks on the \textsf{Drugs} rumor, which had spread the highest. The second most popular claim, \textsf{Weather}, was also the second most fact-checked claim, with 40.3\% of respondents seeing an official source confirming or denying it. Fact-checks for all other claims have been seen by no more than 25\% of respondents on average.

Our results exhibit that fact-checks do not spread at the same rate as the rumors themselves. On average, fewer than half the respondents who have seen rumors have encountered the corresponding fact-checks, as demonstrated by the difference in areas of pink and purple regions in Figure~\ref{fig:countries}. Given that we opt to choose the prevalent rumors on social media, we highlight that only less than half of the people saw the corresponding fact-checked information is quite alarming. Finally, we investigated the relationship between the perceived benefit of sharing fact-checks and exposure to fact-checks. Our results suggest that respondents perceive fact-checks to be more beneficial to their community if they address less commonly seen rumors (Spearman's $\rho$=-0.745, 95\% CI -0.972 -- -0.120).

\subsection*{The Most Vulnerable Regions}

\begin{figure*}[t]
\begin{adjustwidth}{-2.25in}{0in}
\caption{\textbf{Scatter plot and linear relationship between country-level exposure to rumors and ranked GDP per capita.}}
\begin{flushleft}
The x-axis represents the ranked GDP per capita values of countries in our study. Spearman correlation values and their respective significance levels are also presented. The rightmost bottom plot presents the results across all claims. Significance marked as $^*$\textit{p}\textless.05, $^{**}$\textit{p}\textless.01, $^{***}$\textit{p}\textless.001.
\end{flushleft}
    \centering
    \vspace{4mm}
    \includegraphics[width=\linewidth]{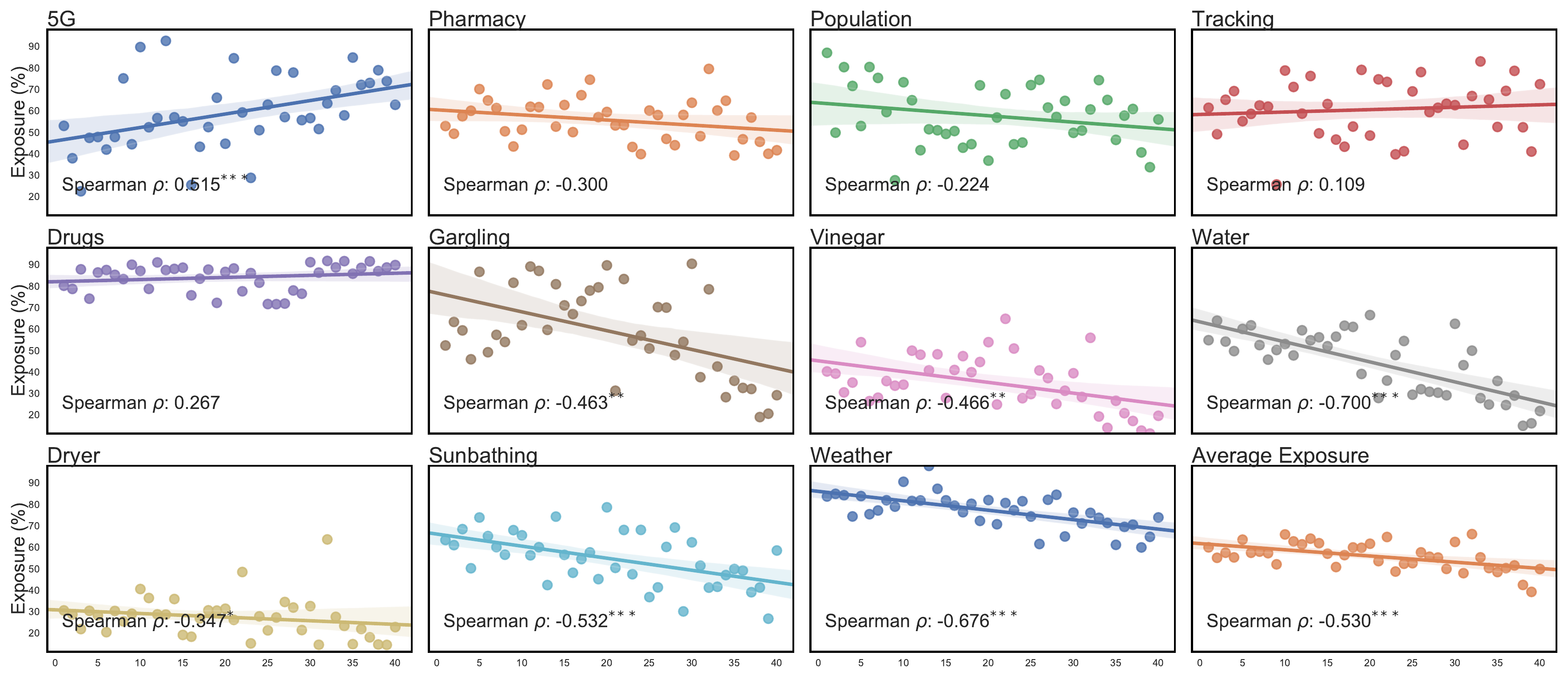}
    \label{fig:exposure}
\end{adjustwidth}
\end{figure*}

We observed that Figure~\ref{fig:countries} seems to suggest a relationship between the infodemic's reach and a country's economic development; developed countries (e.g., Sweden, Spain) appear to have been less exposed to the infodemic than the underdeveloped or developing countries (e.g., Iraq, Nicaragua). Figure~\ref{fig:exposure} shows the level of exposure to each of the eleven rumors by country while ordering countries by their GDP per capita on the x-axis. The choice of GDP per capita is motivated by the fact that it is the most widely used comparative economic indicator. We also assume it as a proxy for health indicators and healthcare infrastructure.

Our results indicate that the \textsf{5G} claim was seen more widely in developed countries than in other regions ($\rho$=0.515, 95\% CI 0.248 -- 0.779). On the other hand, vaccination-related ---  \textsf{Pharmacy}, \textsf{Population}, and \textsf{Tracking} --- and the \textsf{Drugs} claims show no significant difference in exposure between developed and underdeveloped countries. For the remaining claims, we observe a downward trend (i.e., negative correlation), suggesting that disadvantaged nations are more vulnerable to the infodemic. These claims include \textsf{Weather} ($\rho$=-0.676, 95\% CI -0.863 -- -0.485) and DIY measures such as \textsf{Vinegar} ($\rho$=-0.466, 95\% CI -0.777 -- -0.153), \textsf{Sunbath} ($\rho$=-0.532, 95\% CI -0.753 -- -0.309), \textsf{Gargling} ($\rho$=-0.463, 95\% CI -0.786 -- -0.136), \textsf{Water} ($\rho$=-0.700, 95\% CI -0.905 -- -0.491), and \textsf{Dryer} ($\rho$=-0.347, 95\% CI -0.658 -- -0.035).

\subsection*{Believability of Rumors}

\begin{figure*}[t]
\begin{adjustwidth}{-2.25in}{0in}

\caption{\textbf{Relative perceived believability of each rumor addressed in this study.}}
\begin{flushleft}
Country-level z-scores are presented. The countries on the x-axis are in increasing order of GDP per capita.
\end{flushleft}
    \centering
    \vspace{4mm}
    \includegraphics[width=\linewidth]{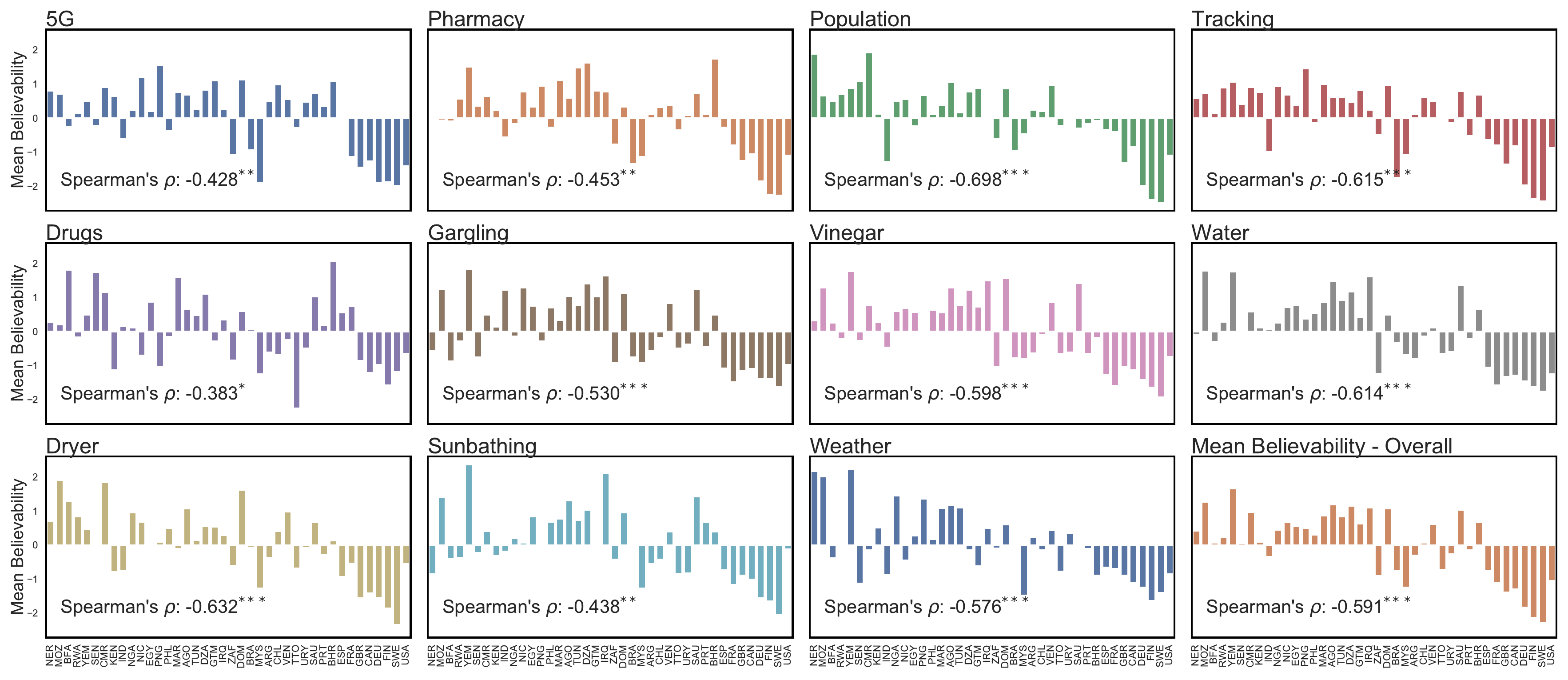}
    \label{fig:belief}

\end{adjustwidth}
\end{figure*}

As a measure of people's perception of these claims, we asked survey respondents to indicate the extent to which they found each claim to be ``believable'' on a 5-point Likert scale. The responses ranged from `Not believable at all' (score of -2) to `Very believable' (score of 2). 

The only claim perceived to be somewhat believable was the \textsf{Drugs} rumor, with a weighted mean value of 0.190 (i.e., borderline believable). All other claims reported a low mean believability (below -0.5), with the \textsf{5G} claim being the least believable claim with a mean value of -1.307. 

To understand what fraction of respondents from each country might be susceptible to believing in misinformation, we dichotomized the reported believability values into positive (i.e., susceptible to believing in a rumor) and negative (i.e., not susceptible). On average, we observe that 22\% of respondents per country are predisposed to believing in rumors. We note the highest misinformation believability (more than 31\%) in Yemen, Algeria, Saudi Arabia, and Tunisia. Swedish and Finnish people seem the least susceptible to the infodemic, with a mere 7.4\% of respondents reporting that rumors are believable.

After determining a country's weighted average for each claim, we calculated a nation's believability z-score. In Figure~\ref{fig:belief}, we present every country's perceived believability (with the countries ranked in increasing order of their GDP per capita) for each rumor. A positive value suggests that a country's population finds a specific claim more believable than an average community, i.e., is more susceptible to believing in it. 

A visual inspection of Figure~\ref{fig:belief} reveals that the most economically developed countries covered by our study show the lowest rates of believability across claims. Conversely, the lower half of the nations ranked by GDP per capita appear to have the highest believability values. These findings suggest that misinformation is likely to be perceived as more believable in economically vulnerable countries.

%% file: content/3_1vacc.tex
\subsection*{Misinformation and Vaccine Acceptance}

After quantifying public exposure to misinformation, we assessed the perceived believability of each rumor (termed \textit{believability} hereafter) and its relationship to willingness to get vaccinated (termed \textit{vaccine acceptance}). Our primary goal was to examine whether extensive exposure to misinformation increases rumor believability and whether such reinforcement further leads to vaccine hesitancy (i.e., a decrease in vaccine acceptance).
 
We analyzed the survey responses through three different regression models (see SI for details). The first model (Model 1) predicts average believability based on one's exposure to misinformation and its corresponding fact-checks. The next two models examine which specific aspects of infodemic are associated with changes in attitudes towards COVID-19 vaccination. For Model 2, we aggregate responses across all claims, while for Model 3, we cluster them into different groups: i) DIY measures, ii) temperature-related (termed Hot\&Co), iii) vaccination conspiracies, and iv) 5G conspiracy. The rationale behind this grouping is to test the influence of distinct notions of misinformation and behavioral measures on vaccine acceptance. Finally, all the presented models control for demographic features and consider other survey responses as independent variables. 
Table~\ref{tab:models} reports, for each model, the average marginal effects (M). 
The regression coefficients are reported in SI Table S3 - S14 for fixed-, mixed-effects, lasso and elastic net models. All models show similar results.

\begin{table}
\begin{adjustwidth}{-2.25in}{0in}
\centering
\caption{\textbf{Summary of regression analysis.}}
\small
\begin{tabular}{{|l+c|c|c|c|}} \hline
{\textbf{Predictors}} & {\textbf{Avg. Believability}} & {\textbf{Vacc. Acceptance}} & {\textbf{Vacc. Acceptance Grouped}}\\

& {\textbf{(Model 1)}} & {\textbf{(Model 2)}} & {\textbf{(Model 3)}}\\ \thickhline

\multicolumn{1}{|l} {\textbf{Control Variables}} & \multicolumn{3}{c|} {} \\ \hline

\quad \textit{Perceived Threat}  & $0.114^{***} (0.095~\mbox{--}~0.133)$ & $0.166^{***} (0.155~\mbox{--}~0.177)$ & $0.147^{***}(0.135~\mbox{--}~0.158)$\\ \hline

\quad \textit{Past Vacc.}  &{--} & $-0.029^{**} (-0.046~\mbox{--}~-0.012)$ & $0.003 (-0.014~$\mbox{--}$~0.020)$\\  \hline

\quad \textit{Past Non-Mandatory Vacc.} &{\mbox{--}} & $0.184^{***} (0.170~\mbox{--}~0.199)$ & $0.171^{***} (0.157~\mbox{--}~0.185)$\\ \hline

\multicolumn{1}{|l} {$\textbf{Independent Variables}$} & \multicolumn{3}{c|} {} \\ \hline

\quad \textit{Exposure}   & $0.098^{***} (0.093~\mbox{--}~0.103)$ &$-0.004^{*} (-0.007~\mbox{--}~-0.001)$ &{\mbox{--}}\\ \hline

\quad \textit{Believability}  &{\mbox{--}}& $-0.128^{***} (-0.137~\mbox{--}~-0.119)$ &{\mbox{--}}\\ \hline

\quad Fact-Checks  & $-0.010^{**} (-0.015~\mbox{--}~-0.004)$ & $0.016^{***} (0.012~\mbox{--}~0.019)$
&{\mbox{--}} \\  \hline

\multicolumn{1}{|l}{\textbf{5G Claims}}  & \multicolumn{3}{c|} {}\\ \hline

\quad \textit{Exposure}  & {\mbox{--}} & {\mbox{--}} & $-0.025^{**} (-0.043~\mbox{--}~-0.007)$\\ \hline

\quad \textit{Believability}  &{\mbox{--}} & {\mbox{--}}& $-0.002 (-0.009~\mbox{--}~0.005)$\\ \hline

\quad \textit{Fact-Checks}  &{\mbox{--}} & {\mbox{--}}& $0.007 (-0.02~\mbox{--}~0.035)$\\ \hline

\multicolumn{1}{|l} {\textbf{DIY Claims}}  & \multicolumn{3}{c|} {}\\ \hline

\quad \textit{Exposure} &{\mbox{--}} & {\mbox{--}} & $0.024^{***} (0.017 -- 0.031)$\\  \hline

\quad \textit{Believability}  &{\mbox{--}} & {\mbox{--}}& $0.001 (-0.010~\mbox{--}~0.012)$\\ \hline

\quad \textit{Fact-Checks}  &{\mbox{--}}&{\mbox{--}}& $0.005 (-0.004~\mbox{--}~0.014)$\\ \hline

\multicolumn{1}{|l}{\textbf{Hot\&Co Claims}} & \multicolumn{3}{c|} {}\\ \hline

\quad \textit{Exposure} &{\mbox{--}}&{\mbox{--}}& $0.010^{*} (0.001~\mbox{--}~0.019)$\\ \hline

\quad \textit{Believability} &{\mbox{--}}&{\mbox{--}}& $0.012^{*} (0.003~\mbox{--}~0.022)$\\ \hline

\quad \textit{Fact-Checks}  &{\mbox{--}}&{\mbox{--}}& $0.009 (-0.001~\mbox{--}~0.020)$\\ \hline

\multicolumn{1}{|l} {\textbf{Vaccination Claims}} & \multicolumn{3}{c|} {} \\ \hline

\quad \textit{Exposure} &{\mbox{--}}&{\mbox{--}}& $-0.035^{***} (-0.042~\mbox{--}~-0.027)$\\ \hline

\quad \textit{Believability} &{\mbox{--}}&{\mbox{--}}& $-0.120^{***} (-0.128~\mbox{--}~-0.113)$\\  \hline

\quad \textit{Fact-Checks}   &{\mbox{--}}&{\mbox{--}}& $0.032^{***} (0.018~\mbox{--}~0.045)$\\ \hline
\end{tabular}
\vspace{2mm}
\begin{flushleft} The table presents the average marginal effects of all main predictors across the three models proposed by the study and their corresponding 95\% confidence intervals. The scale of the variables can be referred to in SI Table S2. Significance marked as $^*p<.05$, $^{**}p\textless.01$, $^{***}p\textless.001$. In addition, we present fixed-, mixed-effects, lasso and elastic net model regression coefficients in SI Table S3 - S14.
\end{flushleft}
\label{tab:models}
\end{adjustwidth}
\end{table}

For Model 1, we find that exposure to misinformation is positively correlated with overall claim believability ({M}=0.098, 95\% CI 0.093 -- 0.103). A higher perceived threat concerning the pandemic is also associated with higher believability ({M}=0.114, 95\% CI 0.095 -- 0.133). Our findings indicate a weak effect of exposure to fact-checks in believability ({M}=--0.010, 95\% CI -0.015 -- -0.004).

Model 2 results, in which claims are not grouped, show that mere exposure to misinformation is not strongly associated with vaccination willingness ({M}=--0.004, 95\% CI -0.007 -- -0.001). However, the perceived believability of false information is associated with vaccine refusal ({M}=-0.128, 95\% CI -0.137 -- -0.119). Although statistically significant, the marginal effect size of overall past-vaccination history is negligibly small, while those who had received a non-mandatory vaccine in the past also report higher vaccine acceptance ({M}=0.184, 95\% 0.170 -- 0.199), as those who perceive the pandemic as more threatening ({M}=0.166, 95\% CI 0.155 -- 0.177).

Our final logistic regression model (Model 3), which groups related claims into distinct types, indicates that increased exposure to vaccine-related misinformation is directly associated with an increased level of vaccination hesitancy ({M}=-0.035, 95\% CI -0.042 -- 0.027). A more substantial association is seen for the reported believability of false vaccination-related claims ({M}=-0.120, 95\% CI -0.128 -- -0.113). Our results also show that increased exposure to fact-checked vaccination-related information is correlated with increased vaccine acceptance ({M}=0.032, 95\% CI 0.018 -- 0.045).

\begin{figure}
\caption{\textbf{Model 3's marginal effects of all predictors and their 95\% confidence intervals.}} 
\begin{flushleft}
Variables are color-coded as per the groups. The horizontal dashed lines indicate \textit{Exposure}, \textit{Believability} and \textit{Fact Checks} for different groups. 
\end{flushleft}
    \centering
    \includegraphics[width=0.8\linewidth]{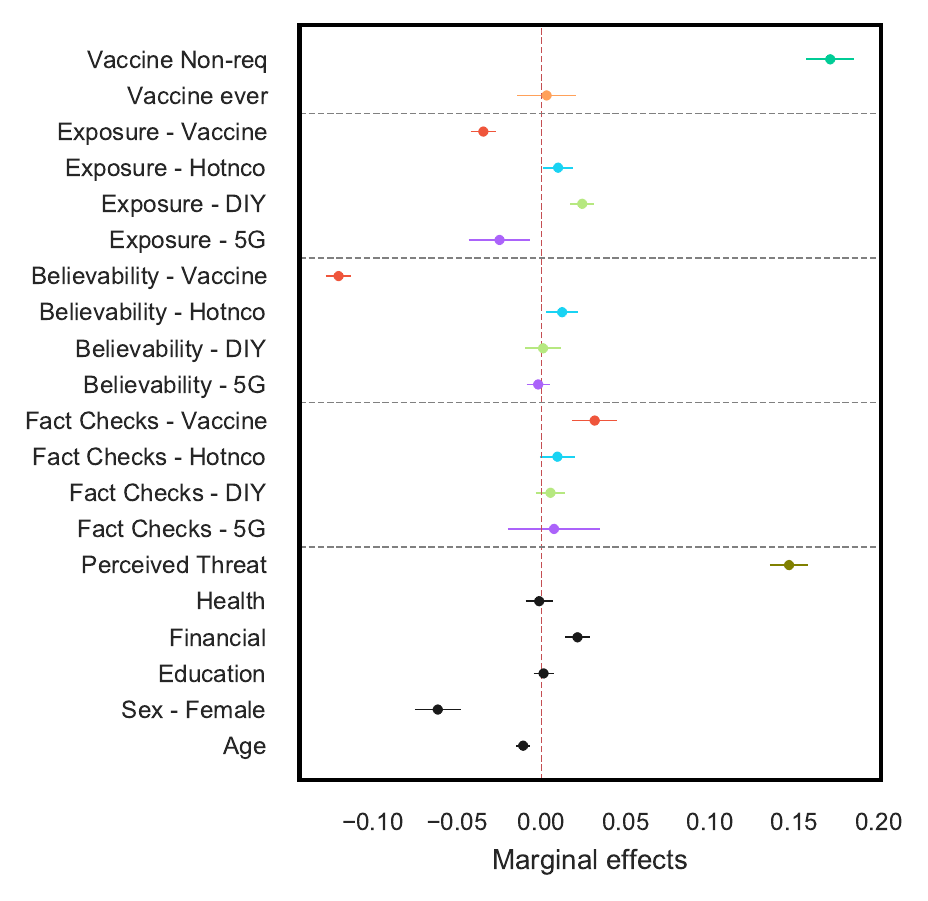}
    \label{fig:plots}
\end{figure}

Aside from the vaccination-related claims, our results suggest that exposure, believability, and fact-checking of other types of misinformation are not strongly associated with vaccine acceptance (see Table~\ref{tab:models}). Additionally, following our Model 2 results, respondents who feel more threatened by the pandemic are more likely to get vaccinated ({M}=0.147, 95\% CI 0.135 -- 0.158). People with previous experience with non-mandatory vaccines also report higher vaccine acceptance ({M}=0.171, 95\% CI 0.157 -- 0.185), whereas general past vaccination has no statistically significant effect. The relative marginal effects for Model 3 can be visualized in Figure~\ref{fig:plots}. 

%% file: content/4discussion.tex
\subsection*{Using Social Media as a Survey Platform}

In the present study, we used the Facebook Advertisement Platform as a recruitment platform for conducting our survey and obtained responses from over 18,000 respondents worldwide. We described our method for collecting demographically diverse survey responses via targeted advertisements for different locations and varying age groups. We also employed a post-stratification weighting scheme (i.e., raking) to correct survey results for non-response and non-coverage. The broad and deep reach of social media and the weighting technique helped us better estimate the infodemic's worldwide reach compared to other commonly used crowdsourcing platforms. 

Additionally, the Facebook Advertisement Platform also is a financially viable choice for conducting global-scale surveys. Our survey, reaching over 50,000 people with 18,407 complete responses, cost US\$8,550. Assuming an identical setting (e.g., a median of 11 minutes to complete the survey), a survey of the same scale could have cost over five times, e.g., US\$43,741 on Prolific~\cite{palan2018prolific} or US\$45,870 on Amazon Mechanical Turk~\cite{paolacci2010running}. Moreover, the demographic reach would be significantly lower with participants mainly from selected regions, e.g., Amazon Mechanical Turk's user base predominantly comprises the US and Indian residents~\cite{ross2010crowdworkers}, and Prolific's workers reside in OECD countries. 
We believe the economic viability and broad geographic reach make social media advertising platforms a feasible survey tool for researchers and practitioners.

\subsection*{Treatment of Local Versus Global Rumors}

Another finding of this paper was the uneven spread of rumors across geographic regions (see Fig.~\ref{fig:countries}). For instance, the top-2 rumors, i.e., \textsf{Drugs} and \textsf{Weather} reached nearly three-fourth of all respondents globally. Besides an inherent appeal of these claims, they might have spread more widely for political reasons. Public figures worldwide downplayed the virus' impact by stating that it would disappear as temperatures started rising~\cite{trump2020weather}. The potential use of existing drugs like hydroxychloroquine, a malaria drug that has not shown any promising result~\cite{horby2020effect}, has been openly promoted as a potential therapy~\cite{bozo2020hydro,trump2020hydro}. These findings exemplify the influence that local public figures can exert on the general public's information, as shown in a prior study on the public narrative of Ibuprofen's possible side-effects on coronavirus patients~\cite{xaudiera2020ibuprofen}.

One of the localized rumors was that saltwater gargling prevents the coronavirus infection (\textsf{Gargling}). This claim had exceptionally high exposure in the Middle East; nearly 90\% of respondents from Saudi Arabia and Iraq reported to have seen the claim compared to only around 20\% of respondents in Sweden and Finland. Although the rumor's premise may be harmless, this claim led to tens of infection in South Korea as some churchgoers continued to congregate after spraying saltwater in each other's mouths~\cite{kh20saltwater}. There have been numerous cases where seemingly harmless misinformation swayed people away from official guidelines (e.g., social distancing, washing hands with soap). In addition to tackling globally popular misinformation, local governments could work together with platforms to further debunk claims with a strong regional foothold.

\subsection*{Algorithmic Prioritization of Fact-Checks}

When it comes to prioritizing claims to debunk first or deciding which fact-checks to disseminate widely, fact-checking organizations need to consider the exposure and believability of claims. A good example to discuss is \textsf{5G}, which was seen by 60\% of all respondents, yet had low believability of, on average, 13\%. This may indicate that relative to its wide dissemination, the potential harms may not be extensive given people are not susceptible to believing it. 

Based on rumor exposure, fact-check exposure, and believability, we propose a heuristic algorithmic prioritization method to decide which fact-checks to disseminate widely first. We propose an estimate of \emph{blind belief} in a rumor as 
\begin{equation}
(\textrm{Rumor Exposure} - \textrm{Fact-Check Exposure}) \times \textrm{(Believability)}
\label{equ:heuristic}
\end{equation}
Using the dichotomized value of believability, we can roughly estimate a proxy for how many people may believe a specific claim. The same idea can be used in deciding which claims to debunk first. 

\begin{figure}
\caption{
        \textbf{Comparison of claims ranked based on (a) heuristic algorithmic prioritization and (b) how currently practiced.}
        }
        \vspace{2mm}
     \centering
     \begin{subfigure}[b]{\linewidth}
     \caption{Blind belief scale (\% of respondents who will likely believe in claims upon exposure, without having access to fact-checks).
        }
        \centering
        \includegraphics[width=.8\linewidth]{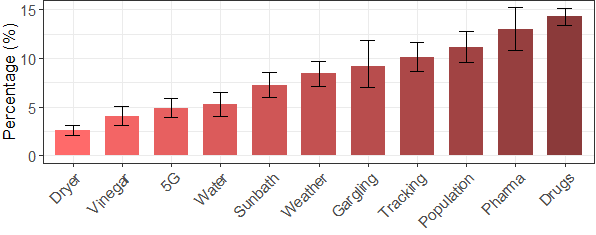}

        \vspace{2.5mm}
        \label{fig:est_belief}
     \end{subfigure}
     \begin{subfigure}[b]{\linewidth}
     \caption{
        Aggregate dissemination percentage of fact-checks.
        }
        \centering
        \includegraphics[width=.8\linewidth]{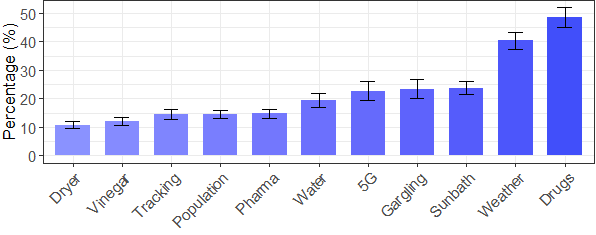}
        \vspace{-5mm}
        
        \label{fig:fact}
     \end{subfigure}
        
        \label{fig:double}
\end{figure}

Figure~\ref{fig:double}(a) shows the estimated percentage of respondents that might believe each claim after encountering it online without having seen a fact-check. \textsf{Drugs} exhibits the highest value, followed by the three vaccination-related claims. Although \textsf{Weather} was the second most seen claim, it is ranked sixth in the blind belief scale. Rumors addressing DIY measures against the disease (e.g., \textsf{Gargling}, \textsf{Sunbath}) suggests that less than 10\% of the population might believe them without access to fact-checks. Likewise, \textsf{5G} is ranked ninth, low if compared to its disproportionate large exposure.  

In contrast, Figure~\ref{fig:double}(b) shows the popularity of fact-checks by the weighted percentage of survey participants who have seen them. Note that \textsf{Drugs} is top-ranked as in the algorithmic suggestion. However, our results suggest that the three vaccination-related fact-checks, which the proposed algorithm recommends to be highly prioritized, have not been widely popular in real campaigns. We also see relatively high dissemination for \textsf{Weather} and \textsf{Sunbath} compared to their blind belief potential.

The observation above implies that fact-checking organizations and social media platforms could use simple online tools to elicit users' perceived believability and identify which claims are more likely to be widely believed. A possibility would be social media platforms presenting prompts to users inquiring whether they find claims believable as soon as they are identified in online networks (e.g., by machine learning methods or reporting functions). Efforts could prioritize those claims that are widely shared \emph{and} perceived to be believable. This prioritization could prove to be incredibly helpful, considering limited manual fact-checking resources and the relative inaccuracy of automated fact-checking models.

\subsection*{Underdeveloped Countries Are More Susceptible to the Infodemic}

Another finding of this study is that economically disadvantaged countries are exposed more to the infodemic than richer nations. Moreover, respondents from nations with lower GDP per capita generally are more susceptible to believing in misinformation upon exposure. This finding is most prominent for claims that propose DIY preventive measures, such as \textsf{Gargling} and \textsf{Water}. It can be linked to the literature which has found that those more economically disadvantaged are more likely to have limited health literacy~\cite{eu2015hliteracy,lorini2018}. Users from these countries have limited access to healthcare~\cite{corscadden2018factors,meyer2013inequities}, which might make them more receptive to non-conventional health behaviors.

Underdeveloped countries seem to be in higher distress during the pandemic~\cite{zar2020challenges}, mainly due to a lack of healthcare infrastructure and limited number of health professionals~\cite{med2015dist}. The spread of the infodemic could increase the burden caused by COVID-19 in these countries, as health information inequalities are known to widen global health disparities~\cite{hudspeth2017health}. 

A Pew Research Center study has found that social media use is continuously rising in underdeveloped and developing countries~\cite{poushter2018social}. This observation is two-fold. Although people might be exposed to more misinformation as they go online, this also creates opportunities for online dissemination of accurate and debunking information. Hence, our finding underlines the importance of fact-checking platforms in propagating correct information \emph{before} rumors spread in vulnerable countries. Preparing reactive fact-checks alone might not be enough if their spreading potential is smaller than rumors' and if rumors are considered believable. Social media platforms could be used to facilitate the dissemination of such preemptive fact-checks.

\subsection*{Vaccination-Related Rumors Are Popular Worldwide}

Vaccination-related claims show no significant difference in exposure between developed and underdeveloped countries. More worryingly, our results suggest that these claims are widely shared online, reporting up to 60\% exposure across the countries covered by our study. Our proposed estimate of public belief in rumors presents these claims as prospects for high belief among the world population. This is a concern for global health as research has found that information delivered through social media can cause COVID-19 vaccine hesitancy~\cite{puri2020social}.

Given the importance of vaccination in the control of the pandemic~\cite{draulans2020finally} and the strong influence that anti-vaccination movements exert in online communities~\cite{johnson2020online}, we highlight the importance of fighting this type of misinformation, particularly now that coronavirus vaccines are being rolled out. Some online platforms have taken proactive stances on this topic and banned misinformation about COVID-19 vaccines~\cite{vergeytvacc,vergefbvacc}. Our findings highlight the importance of these efforts, and we urge other social platforms to adopt a similar stance on the topic.

\subsection*{Infodemic and Vaccine Hesitancy}

Our analysis reveals that exposure to misinformation influences vaccination decisions. Interestingly, false information's perceived believability is a much more decisive factor in vaccine acceptance than mere exposure. Susceptibility to believing in misinformation, and consequent belief in an unconfirmed piece of information, could have critical implications on public health behaviors. On the other hand, increased exposure to fact-checked information is associated with a more positive attitude towards the coronavirus vaccine. Although the adverse impact of perceived believability of misinformation in vaccine acceptance is more pronounced than that of fact-checked information, its positive effect highlights the importance of concerted efforts for disseminating accurate and debunking information to the public.

As the claims addressed in this study cover various aspects of the infodemic, we also studied whether different misinformation categories have varying effects on people's vaccination tendencies. Although increased exposure to vaccination-related false information and associated believability negatively affects vaccine acceptance, our results suggest marginally adverse or even positive effects for other types of misinformation (see Fig~\ref{fig:plots}).

These conflicting and marginal results indicate that misinformation not directly addressing vaccination might not be associated with vaccine refusal. For instance, those exposed to more DIY-related claims show higher rates of vaccine acceptance; people adhering to various behavioral measures for their safety might also feel more threatened about the coronavirus and thus may be more willing to accept a vaccine. Another hypothesis is that people interested in personal health and well-being, i.e., arguably more likely to have seen DIY rumors, are active followers of coronavirus-related information to protect themselves from infection. The opposite effect was observed for the \textsf{5G} rumor; people who have seen this conspiracy theory might also have been exposed to vaccination-related conspiracies~\cite{5g20guardian} and hence show higher vaccine refusal rates.

%% file: content/5conclusion.tex
To understand how the COVID-19 infodemic has affected different countries worldwide, we conducted a large-scale survey to quantify public exposure to a wide range of coronavirus-related misinformation and fact-checks. Additionally, we assessed the extent to which people's belief in misinformation negatively influences their acceptance of the coronavirus vaccine.  
%
All forty countries examined showed extensively higher exposure to rumors than to their respective fact-checks. Most importantly, our findings indicate that the infodemic could disproportionately hit economically disadvantaged countries the hardest. These vulnerable countries have higher rates of exposure to coronavirus-related rumors and found claims more believable than developed nations' residents.

Our study indicates that misinformation, particularly to what extent people are open to believing in it, negatively influences their acceptance of the coronavirus vaccine. A more fine-grained analysis revealed how vaccination-related claims could contribute to vaccination hesitancy, while other false information does not seem to influence these decisions. Worryingly, our findings indicate that the positive effect of fact-checks on vaccine acceptance is less pronounced than the extent to which the population is susceptible to believing in misinformation.

There are however, several limittaions that might be associated with this work. Although we have designed our study to cover a wide range of coronavirus-related misinformation topics, our list is not comprehensive of the whole infodemic. Future work should address a more extensive list of rumors which were not covered in the current study. We have also conducted our study in a month-long time window. The infodemic is under constant mutation, and future studies should also address the temporal aspect of misinformation. This also means that in retrospect, the ever-evolving nature of pandemic and the associated infodemic might also introduce some asymmetries in the survey design. Although we have adopted post-stratification methods to compensate for non-respondents and non-coverage, our results might not be generalizable to those countries with smaller sample sizes. Additionally, our weighting technique, i.e., raking, might be associated with issues like non-converge under some conditions~\cite{ireland1968contingency, Battaglia2009Practical}.

We recruited our respondents through social media to cover a wide range of respondents from different world regions and adopted weighting methods to compensate for non-respondents, but our results are not strictly representative of the world population. 
For instance, we have focused our efforts on economically undeveloped countries (e.g., in Africa), as previous work indicates that developing countries are more vulnerable to communicable diseases~\cite{ncd19book}. Hence, our respondents do not cover the majority of other countries. Facebook users could also be more susceptible to being exposed to rumors, as false information is rapidly disseminated online. Moreover, we maximized the reach of our survey by translating it into some of the most widely spoken languages, but we did not cover Chinese, Indic, and Slavic languages.

Our results are based on self-reported values, which could be influenced by social desirability biases~\cite{fisher1993social}. An influencing factor could also be the terminology used in the survey. For future work, it could be mitigated by a more careful survey design and the use of clearer phrasing. Nevertheless, the association between perceived believability (particularly of vaccine-related rumors) and vaccine hesitancy is highly significant across a wide range of models. Additionally, we underline that previous studies have not observed social desirability biases in estimates of compliance with COVID-19 regulations~\cite{larsen2020survey}. Hence, our main findings associating rumor exposure and believability with vaccine hesitancy should be consistent with Facebook users' views.

It is also important to consider a possible limitation concerning self-selection biases; participants who chose to take part in the study by clicking on its advertisement might have been particularly interested in the pandemic. Nevertheless, we highlight that previous studies have found no major bias in Facebook samples compared to traditionally administered surveys~\cite{kalimeri2019evaluation}, particularly if correction factors, such as post-stratification weights, are used~\cite{ribeiro2020biased}.

The fact that we quantified infodemic spread and its association with vaccine hesitancy at a global scale makes our work truly unique. We present our analysis with individual responses from $40$ countries of world, covering the continents of Asia, Africa, Europe, and the Americas, with translations in widely spoken local languages. This work is also a distinction from published research on the matter that is largely Anglo-centric.

We defended a proactive stance in disseminating accurate information, flagging suspicious content before rumors are widely spread and believed. We also highlighted the importance of local efforts in fighting the infodemic as claims might be constrained to specific regions. Given the importance of perceived believability in overall belief in claims regardless of their reach, we discussed how social media platforms could elicit users' perceived believability of rumors to prioritize which misinformation should be quickly addressed. In addition, we proposed a heuristic prioritization method based on the reach of rumors, fact-checks, and perceived believability to assist decisions of which rumors to prioritize in the fight against the infodemic. 

Finally, our findings suggest that accurate information promulgating public awareness about the disease's risks and side effects is important for widespread vaccine acceptance. We underlined the importance of debunking falsehoods and spreading truths about vaccination given the public susceptibility to and widespread reach of vaccination-related rumors, particularly once coronavirus vaccines are distributed worldwide. Additionally, our study indicates that economically vulnerable countries are more susceptible to infodemic. Considering these nations' more precarious healthcare infrastructure, we propose that public organizations and social media platforms prioritize underdeveloped and developing countries to fight against misinformation. Social media is continuously growing in these nations, and platforms should play a key role in suppressing rumors and disseminating facts. We hope findings from this work contribute to public and policy decisions and enable the authorities and the stakeholders to be better equipped for responding better to future misinformation spreading. 

%% file: content/6si.tex

\textbf{S1 - S14 Tables}

\begin{table}[h!]
\begin{adjustwidth}{-2.25in}{0in}
\centering
\footnotesize
\caption{\textbf{Demographic distribution of study participants by country.}}
\begin{tabular}{l|r|r|rrrrrr|rrr}
\toprule
& & Gender & \multicolumn{6}{c|}{Age} & \multicolumn{3}{c}{Education} \\
\hline
Country & N & Female & 18-24 & 25-34 & 35-44 & 45-54 & 55-64 & 65+ & HS or Lower & BS or Assoc. Degree & Grad. Degree \\ 
  AGO & 355 & 54.9\% & 29\% & 22.8\% & 18.9\% & 11.5\% & 13\% & 4.8\% & 50.7\% & 33.8\% & 15.5\% \\ 
  ARG & 1086 & 25.5\% & 8.1\% & 7.1\% & 21\% & 15.8\% & 29.9\% & 18\% & 48.2\% & 9.9\% & 42\% \\ 
  BFA &  75 & 77.3\% & 16\% & 21.3\% & 22.7\% & 12\% & 18.7\% & 9.3\% & 24\% & 32\% & 44\% \\ 
  BHR &  31 & 71\% & 9.7\% & 29\% & 22.6\% & 12.9\% & 12.9\% & 12.9\% & 16.1\% & 71\% & 12.9\% \\ 
  BRA & 701 & 23.4\% & 29\% & 11.4\% & 22.8\% & 9.7\% & 16.5\% & 10.6\% & 54.1\% & 30\% & 16\% \\ 
  CAN & 871 & 36.4\% & 10.8\% & 7.9\% & 13.2\% & 11.1\% & 27.2\% & 29.7\% & 48.7\% & 34.8\% & 16.5\% \\ 
  CHL & 569 & 19\% & 15.6\% & 7.4\% & 26\% & 17\% & 21.1\% & 12.8\% & 52.4\% & 8.8\% & 38.8\% \\ 
  CMR &  93 & 58.1\% & 10.8\% & 23.7\% & 30.1\% & 11.8\% & 20.4\% & 3.2\% & 25.8\% & 24.7\% & 49.5\% \\ 
  DEU & 631 & 41\% & 31.7\% & 9.4\% & 14.4\% & 17.4\% & 16.6\% & 10.5\% & 43.6\% & 26.9\% & 29.5\% \\ 
  DOM & 481 & 24.3\% & 21.6\% & 24.5\% & 23.5\% & 13.5\% & 11.2\% & 5.6\% & 36.8\% & 12.5\% & 50.7\% \\ 
  DZA & 1061 & 66.6\% & 22.8\% & 33.3\% & 29.9\% & 10.7\% & 3\% & 0.3\% & 19.2\% & 50.9\% & 29.9\% \\ 
  EGY & 639 & 60.1\% & 33.6\% & 26\% & 23.6\% & 10.6\% & 4.2\% & 1.9\% & 30.4\% & 57\% & 12.7\% \\ 
  ESP & 363 & 35.8\% & 16.5\% & 11.3\% & 21.8\% & 16.8\% & 20.7\% & 12.9\% & 32\% & 24\% & 44.1\% \\ 
  FIN & 418 & 40\% & 28.9\% & 15.6\% & 19.1\% & 17.7\% & 9.3\% & 9.3\% & 45.2\% & 30.6\% & 24.2\% \\ 
  FRA & 728 & 37.5\% & 14.7\% & 11.5\% & 19.4\% & 13.7\% & 19.9\% & 20.7\% & 36.4\% & 19.1\% & 44.5\% \\ 
  GBR & 614 & 30.6\% & 14.5\% & 7.5\% & 21.8\% & 6.8\% & 16.8\% & 32.6\% & 47.9\% & 31.4\% & 20.7\% \\ 
  GTM & 681 & 37.9\% & 21.7\% & 18.6\% & 23.1\% & 16.4\% & 13.1\% & 7\% & 46.1\% & 14.7\% & 39.2\% \\ 
  IND & 229 & 63.3\% & 24\% & 8.3\% & 11.8\% & 16.2\% & 14.8\% & 24.9\% & 21\% & 29.3\% & 49.8\% \\ 
  IRQ & 883 & 75.9\% & 22.7\% & 32.2\% & 22.2\% & 16\% & 4.9\% & 2.2\% & 29\% & 61.7\% & 9.3\% \\ 
  KEN & 207 & 51.7\% & 14.5\% & 31.4\% & 29.5\% & 15.5\% & 6.8\% & 2.4\% & 42.5\% & 38.6\% & 18.8\% \\ 
  MAR & 454 & 70.7\% & 28\% & 32.6\% & 23.3\% & 8.1\% & 5.1\% & 2.9\% & 26.9\% & 46.9\% & 26.2\% \\ 
  MOZ & 414 & 57.7\% & 19.1\% & 31.2\% & 20.5\% & 13.5\% & 12.3\% & 3.4\% & 42.5\% & 46.1\% & 11.4\% \\ 
  MYS & 241 & 41.9\% & 32.8\% & 21.2\% & 15.8\% & 10.8\% & 14.1\% & 5.4\% & 44.8\% & 36.1\% & 19.1\% \\ 
  NER &  60 & 85\% & 13.3\% & 25\% & 20\% & 31.7\% & 10\% & 0\% & 23.3\% & 25\% & 51.7\% \\ 
  NGA & 300 & 66.7\% & 16.7\% & 30.3\% & 27.7\% & 14.3\% & 9.7\% & 1.3\% & 21\% & 52.7\% & 26.3\% \\ 
  NIC & 528 & 40.2\% & 25.8\% & 19.7\% & 17.6\% & 14.8\% & 13.6\% & 8.5\% & 34.1\% & 8.7\% & 57.2\% \\ 
  PHL & 621 & 51\% & 24.5\% & 12.9\% & 19.6\% & 11.8\% & 17.2\% & 14\% & 32\% & 49.9\% & 18\% \\ 
  PNG &  51 & 78.4\% & 9.8\% & 5.9\% & 31.4\% & 19.6\% & 21.6\% & 11.8\% & 47.1\% & 35.3\% & 17.6\% \\ 
  PRT & 520 & 28.7\% & 28.7\% & 9\% & 26.9\% & 12.9\% & 12.1\% & 10.4\% & 49.8\% & 37.7\% & 12.5\% \\ 
  RWA & 106 & 72.6\% & 6.6\% & 21.7\% & 30.2\% & 25.5\% & 12.3\% & 3.8\% & 12.3\% & 60.4\% & 27.4\% \\ 
  SAU &  76 & 52.6\% & 6.6\% & 25\% & 30.3\% & 19.7\% & 14.5\% & 3.9\% & 22.4\% & 53.9\% & 23.7\% \\ 
  SEN & 234 & 50.4\% & 10.3\% & 18.4\% & 17.1\% & 14.5\% & 23.5\% & 16.2\% & 28.6\% & 19.7\% & 51.7\% \\ 
  SWE & 779 & 40.9\% & 24.4\% & 15.5\% & 22.1\% & 14.8\% & 13.1\% & 10.1\% & 40.8\% & 30.7\% & 28.5\% \\ 
  TTO & 353 & 32\% & 11.3\% & 12.2\% & 24.1\% & 20.1\% & 19.3\% & 13\% & 39.9\% & 40.8\% & 19.3\% \\ 
  TUN & 436 & 54.8\% & 29.6\% & 27.1\% & 25.7\% & 12.4\% & 3.4\% & 1.8\% & 23.9\% & 47.5\% & 28.7\% \\ 
  URY & 542 & 23.1\% & 11.3\% & 6.5\% & 18.8\% & 18.6\% & 26.6\% & 18.3\% & 47\% & 19.7\% & 33.2\% \\ 
  USA & 378 & 40.5\% & 14.8\% & 8.2\% & 12.7\% & 10.8\% & 22\% & 31.5\% & 37.3\% & 38.6\% & 24.1\% \\ 
  VEN & 459 & 35.1\% & 10\% & 10.5\% & 15.3\% & 18.7\% & 25.3\% & 20.3\% & 27.5\% & 13.9\% & 58.6\% \\ 
  YEM & 342 & 68.1\% & 27.2\% & 32.7\% & 25.4\% & 11.4\% & 2.6\% & 0.6\% & 28.4\% & 60.5\% & 11.1\% \\ 
  ZAF & 704 & 21.6\% & 9.7\% & 8.9\% & 18.9\% & 17.5\% & 26.3\% & 18.8\% & 53.7\% & 27.1\% & 19.2\% \\ 
   \bottomrule
\end{tabular}
   \label{tab:dem}
\end{adjustwidth}
\end{table}



\clearpage


\begin{table}[h!]
\centering
\caption{\textbf{Coding rules used for regression analysis of believability and vaccine acceptance.}}
\begin{tabular}{|l+l|l|l|}
\hline
\textbf{Variable} & \textbf{Values}  & \textbf{Usage} \\ \hline
\thickhline
\textit{Age}  & 0~--~5 & Coded Values\\ \hline
\textit{Sex}  & 0 \& 1 & Coded Values\\ \hline
\textit{Education}  & 0~--~4 & Coded Values\\ \hline
\textit{Financial}  & -2~--~2 &  Coded Values\\ \hline
\textit{Vaccine History}  & 0 \& 1 & Factor\\ \hline
\textit{Perceived Threat}  & 0~--~3 &  Mean\\ \hline
\textit{Exposure}  & 0 \& 1 &  Count\\ \hline
\textit{Fact-Checks} & 0 \& 1 & Count\\ \hline
\textit{Believability}  & -2~--~2 & Mean\\ \hline
\textit{Vaccine Decision} & 0 \& 1 & Factor\\ \hline
\end{tabular}

\label{tab:codes}
\end{table}




%% file: content/model_results.tex

\begin{table}[h!] \centering 
\caption{\textbf{Model 1 regression results. Average believability is predicted from exposure to false claims and their respective fact-checks. Standard errors are presented between parenthesis. Significance marked as $^{*}$ P$<$0.1; $^{**}$ P$<$0.05; $^{***}$ P$<$0.01.}}
\small
\begin{tabular}{@{\extracolsep{5pt}}lc} 
\toprule
 & \multicolumn{1}{c}{\textit{Dependent Variable:}} \\ 
\cline{2-2} 
\\[-1.8ex] & Average Believability \\ 
\hline \\[-1.8ex] 
 Constant & $-$1.399$^{***}$ \\ 
  & (0.032) \\ 
 Sex & $-$0.034$^{***}$ \\ 
  & (0.011) \\ 
 Age & $-$0.009$^{**}$ \\ 
  & (0.003) \\ 
 Education & $-$0.027$^{***}$ \\ 
  & (0.005) \\ 
 Financial Status & $-$0.111$^{***}$ \\ 
  & (0.006) \\ 
 Health Status & 0.071$^{***}$ \\ 
  & (0.007) \\ 
 Perceived Threat & 0.114$^{***}$ \\ 
  & (0.010) \\ 
  \hline
 Fact-Checks & $-$0.068$^{***}$ \\ 
  & (0.007) \\ 
 Exposure & 0.075$^{***}$ \\ 
  & (0.003) \\ 
 Exposure $\times$ Fact-Checks & 0.010$^{***}$ \\ 
  & (0.001) \\ 
\hline \\[-1.8ex] 
Observations & 18,314 \\ 
$R^2$ & 0.129 \\
\bottomrule
\end{tabular} 
\end{table}

\begin{table}[h!] \centering 
  \caption{\textbf{Model 1 mixed-regression results. Average believability is predicted from exposure to false claims and their respective fact-checks. We include the respondent's residence country as a random effect. Standard errors are presented between parenthesis. Significance marked as $^{*}$ P$<$0.1; $^{**}$ P$<$0.05; $^{***}$ P$<$0.01.}}
  \small
\begin{tabular}{@{\extracolsep{5pt}}lc} 
\\[-1.8ex]\hline 
\hline \\[-1.8ex] 
 & \multicolumn{1}{c}{\textit{Dependent Variable:}} \\ 
\cline{2-2} 
\\[-1.8ex] & Average Believability \\ 
\hline \\[-1.8ex] 
 Constant & $-$0.774$^{***}$ \\ 
  & (0.061) \\ 
 Sex & 0.077$^{***}$ \\ 
  & (0.011) \\ 
 Age & 0.032$^{***}$ \\ 
  & (0.003) \\ 
 Education & $-$0.058$^{***}$ \\ 
  & (0.005) \\ 
 Financial Status & $-$0.091$^{***}$ \\ 
  & (0.006) \\ 
 Health Status & 0.018$^{***}$ \\ 
  & (0.007) \\ 
 Perceived Threat & $-$0.085$^{***}$ \\ 
  & (0.010) \\ 
  \hline
 Fact-Checks & $-$0.067$^{***}$ \\ 
  & (0.007) \\ 
 Exposure & 0.051$^{***}$ \\ 
  & (0.003) \\ 
 Exposure $\times$ Fact-Checks & 0.010$^{***}$ \\ 
  & (0.001) \\ 
\hline \\[-1.8ex] 
Country-Level Random Effects & Yes \\
\hline \\[-1.8ex] 
Observations & 18,314 \\ 
$R^2$ & 0.247 \\
\bottomrule
\end{tabular} 
\end{table} 

\begin{table}[h!] \centering 
\caption{\textbf{Model 1 elastic regression results. Average believability is predicted from exposure to false claims and their respective fact-checks.} }
\small
\begin{tabular}{@{\extracolsep{5pt}}lc} 
\\[-1.8ex]\hline 
\hline \\[-1.8ex] 
 & \multicolumn{1}{c}{\textit{Dependent Variable:}} \\ 
\cline{2-2} 
\\[-1.8ex] & Average Believability \\ 
\hline \\[-1.8ex] 
 Constant & $-$1.096 \\ 
 Sex - Male & 0.039 \\ 
 Sex - Female & $-$0.002 \\ 
 Age & $-$0.006 \\ 
 Education & $-$0.014 \\ 
 Financial Status & $-$0.058 \\ 
 Health Status & 0.043 \\ 
 Perceived Threat & 0.103 \\ 
  \hline
 Fact-Checks & $-$0.042 \\ 
 Exposure & 0.049 \\ 
 Exposure $\times$ Fact-Checks & 0.007 \\ 
\hline \\[-1.8ex] 
Observations & 18,314 \\ 
\bottomrule
\end{tabular} 
\end{table}

\begin{table}[h!] \centering 
  \caption{\textbf{Model 1 lasso regression results. Average believability is predicted from exposure to false claims and their respective fact-checks.} }
  \small
\begin{tabular}{@{\extracolsep{5pt}}lc} 
\\[-1.8ex]\hline 
\hline \\[-1.8ex] 
 & \multicolumn{1}{c}{\textit{Dependent Variable:}} \\ 
\cline{2-2} 
\\[-1.8ex] & Average Believability \\ 
\hline \\[-1.8ex] 
 Constant & $-$1.100 \\ 
 Sex - Male & 0.040 \\ 
 Sex - Female & 0.000 \\ 
 Age & $-$0.006 \\ 
 Education & $-$0.013 \\ 
 Financial Status & $-$0.058 \\ 
 Health Status & 0.042 \\ 
 Perceived Threat & 0.102 \\ 
  \hline
 Fact-Checks & $-$0.039 \\ 
 Exposure & 0.050 \\ 
 Exposure $\times$ Fact-Checks & 0.006 \\ 
\hline \\[-1.8ex] 
Observations & 18,314 \\ 
\bottomrule
\end{tabular} 
\end{table}


\begin{table}[h!] \centering 
  \caption{\textbf{Model 2 regression results. Each respondents' vaccine decision is predicted based on exposure to false claims and their fact-checks, believability, perceived threat, and vaccination history. Standard errors are presented between parenthesis. Significance marked as $^{*}$ P$<$0.1; $^{**}$ P$<$0.05; $^{***}$ P$<$0.01.}}
 \small
\begin{tabular}{@{\extracolsep{5pt}}lc} 
\toprule
 & \multicolumn{1}{c}{\textit{Dependent Variable:}} \\ 
\cline{2-2} 
\\[-1.8ex] & Vaccine Acceptance \\ 
\hline \\[-1.8ex] 
 Constant & $-$2.067$^{***}$ \\ 
  & (0.101) \\ 
 Sex & $-$0.307$^{***}$ \\ 
  & (0.033) \\ 
 Age & $-$0.051$^{***}$ \\ 
  & (0.010) \\ 
 Education & 0.019 \\ 
  & (0.014) \\ 
 Financial Status & 0.145$^{***}$ \\ 
  & (0.018) \\ 
 Health Status & $-$0.007 \\ 
  & (0.020) \\ 
 Past Vaccination & $-$0.195$^{***}$ \\ 
  & (0.046) \\ 
 Past Non-Mandatory Vaccination & 0.737$^{***}$ \\ 
  & (0.073) \\ 
  Past Vaccination $\times$ Past Non-Mandatory Vaccination & 0.129 \\ 
  & (0.082) \\ 
 Perceived Threat & 0.775$^{***}$ \\ 
  & (0.029) \\ 
  \hline
 Fact-Checks & 0.061$^{***}$ \\ 
  & (0.021) \\ 
 Exposure & $-$0.023$^{**}$ \\ 
  & (0.009) \\ 
Average Believability & $-$0.598$^{***}$ \\ 
  & (0.023) \\ 
 Exposure $\times$ Fact-Checks & 0.002 \\ 
  & (0.003) \\ 
\hline \\[-1.8ex] 
Observations & 18,314 \\ 
$R^2$ & 0.132 \\
\bottomrule
\end{tabular} 
\end{table} 


\begin{table}[h!] \centering 
  \caption{\textbf{Model 2 mixed-regression results. Each respondents' vaccine decision is predicted based on exposure to false claims and their fact-checks, believability, perceived threat, and vaccination history. We include the respondent's residence country as a random effect. Standard errors are presented between parenthesis. Significance marked as $^{*}$ P$<$0.1; $^{**}$ P$<$0.05; $^{***}$ P$<$0.01.}}
  \small
\begin{tabular}{@{\extracolsep{5pt}}lc} 
\\[-1.8ex]\hline 
\hline \\[-1.8ex] 
 & \multicolumn{1}{c}{\textit{Dependent Variable:}} \\ 
\cline{2-2} 
\\[-1.8ex] & Vaccine Acceptance \\ 
\hline \\[-1.8ex] 
 Constant & $-$1.859$^{***}$ \\ 
  & (0.134) \\ 
 Sex & $-$0.298$^{***}$ \\ 
  & (0.035) \\ 
 Age & $-$0.025$^{**}$ \\ 
  & (0.011) \\ 
 Education & 0.008 \\ 
  & (0.015) \\ 
 Financial Status & 0.151$^{***}$ \\ 
  & (0.019) \\ 
 Health Status & $-$0.064$^{***}$ \\ 
  & (0.021) \\ 
 Past Vaccination & $-$0.155$^{***}$ \\ 
  & (0.052) \\ 
 Past Non-Mandatory Vaccination & 0.611$^{***}$ \\ 
  & (0.076) \\ 
 Past Vaccination $\times$ Past Non-Mandatory Vaccination & 0.273$^{***}$ \\ 
  & (0.085) \\ 
 Perceived Threat & 0.643$^{***}$ \\ 
  & (0.032) \\ 
  \hline
 Fact-Checks & 0.050$^{**}$ \\ 
  & (0.022) \\ 
 Exposure & $-$0.028$^{***}$ \\ 
  & (0.009) \\ 
  Average Believability & $-$0.689$^{***}$ \\ 
  & (0.025) \\ 
 Exposure $\times$ Fact-Checks & 0.003 \\ 
  & (0.003) \\ 
\hline \\[-1.8ex] 
Country-Level Random Effects & Yes \\
\hline \\[-1.8ex] 
Observations & 18,314 \\ 
$R^2$ & 0.169\\
\bottomrule
\end{tabular} 
\end{table} 

\begin{table}[h!] \centering 
  \caption{\textbf{Model 2 elastic regression results. Each respondents' vaccine decision is predicted based on exposure to false claims and their fact-checks, believability, perceived threat, and vaccination history.} }
  \small
\begin{tabular}{@{\extracolsep{5pt}}lc} 
\toprule
 & \multicolumn{1}{c}{\textit{Dependent Variable:}} \\ 
\cline{2-2} 
\\[-1.8ex] & Vaccine Acceptance \\ 
\hline \\[-1.8ex] 
 Constant & $-$2.067 \\ 
 Sex - Male & 0.148 \\ 
 Sex - Female & $-$0.133 \\ 
 Age & $-$0.043 \\ 
 Education & 0.015 \\ 
 Financial Status & 0.127 \\ 
 Health Status & $-$0.001 \\ 
 Past Vaccination & $-$0.173 \\ 
 Past Non-Mandatory Vaccination & 0.650 \\ 
  Past Vaccination $\times$ Past Non-Mandatory Vaccination & 0.181 \\ 
 Perceived Threat & 0.712 \\ 
  \hline
 Fact-Checks & 0.056 \\ 
 Exposure & $-$0.018 \\ 
Average Believability & $-$0.555 \\ 
 Exposure $\times$ Fact-Checks & 0.002 \\ 
\hline \\[-1.8ex] 
Observations & 18,314 \\ 
\bottomrule
\end{tabular} 
\end{table} 

\begin{table}[h!] \centering 
  \caption{\textbf{Model 2 lasso regression results. Each respondents' vaccine decision is predicted based on exposure to false claims and their fact-checks, believability, perceived threat, and vaccination history.} }
  \small
\begin{tabular}{@{\extracolsep{5pt}}lc} 
\toprule
 & \multicolumn{1}{c}{\textit{Dependent Variable:}} \\ 
\cline{2-2} 
\\[-1.8ex] & Vaccine Acceptance \\ 
\hline \\[-1.8ex] 
 Constant & $-$2.370 \\ 
 Sex - Male & 0.296 \\ 
 Sex - Female & 0.000 \\ 
 Age & $-$0.047 \\ 
 Education & 0.015 \\ 
 Financial Status & 0.137 \\ 
 Health Status & 0.000 \\ 
 Past Vaccination & $-$0.174 \\ 
 Past Non-Mandatory Vaccination & 0.745 \\ 
  Past Vaccination $\times$ Past Non-Mandatory Vaccination & 0.103 \\ 
 Perceived Threat & 0.761 \\ 
  \hline
 Fact-Checks & 0.068 \\ 
 Exposure & $-$0.017 \\ 
Average Believability & $-$0.591 \\ 
 Exposure $\times$ Fact-Checks & 0.001 \\ 
\hline \\[-1.8ex] 
Observations & 18,314 \\ 
\bottomrule
\end{tabular} 
\end{table}


\begin{table}[h!] \centering
 \caption{\textbf{Model 3 regression results. Each respondents' vaccine decision is predicted based on exposure to false claims and their fact-checks, believability, perceived threat, and vaccination history. Claims are grouped into their respective topic. Standard errors are presented between parenthesis. Significance marked as $^{*}$ P$<$0.1; $^{**}$ P$<$0.05; $^{***}$ P$<$0.01.} }
 \footnotesize
\begin{tabular}{@{\extracolsep{5pt}}lc} 
\toprule
 & \multicolumn{1}{c}{\textit{Dependent Variable:}} \\ 
\cline{2-2} 
\\[-1.8ex] & Vaccine Acceptance \\ 
\hline \\[-1.8ex] 
 Constant & $-$1.997$^{***}$ \\ 
  & (0.104) \\ 
 Sex & $-$0.304$^{***}$ \\ 
  & (0.035) \\ 
 Age & $-$0.054$^{***}$ \\ 
  & (0.010) \\ 
 Education & 0.006 \\ 
  & (0.015) \\ 
 Financial Status & 0.105$^{***}$ \\ 
  & (0.019) \\ 
 Health Status & $-$0.007 \\ 
  & (0.020) \\ 
 Past Vaccination & $-$0.040 \\ 
  & (0.048) \\ 
 Past Non-Mandatory Vaccination & 0.730$^{***}$ \\ 
  & (0.075) \\ 
 Past Vaccination $\times$ Past Non-Mandatory Vaccination & 0.117 \\ 
  & (0.084) \\ 
 Perceived Threat & 0.724$^{***}$ \\ 
  & (0.030) \\ 
  \hline
 Exposure to Vaccination-Related Claims & $-$0.164$^{***}$ \\ 
  & (0.019) \\ 
 Fact-Checks of Vaccination-Related Claims & 0.182$^{**}$ \\ 
  & (0.073) \\ 
 Average Believability of Vaccination-Related Claims & $-$0.594$^{***}$ \\ 
  & (0.021) \\ 
   Exposure $\times$ Fact-Checks - Vaccination-Related Claims & $-$0.016 \\ 
  & (0.028) \\ 
  \hline
 Exposure to DIY Claims & 0.105$^{***}$ \\ 
  & (0.021) \\ 
 Fact-Checks of DIY Claims & $-$0.004 \\ 
  & (0.045) \\ 
 Average Believability of DIY Claims & 0.005 \\ 
  & (0.027) \\ 
 Exposure $\times$ Fact-Checks - DIY Claims & 0.014 \\ 
  & (0.015) \\ 
  \hline
 Exposure to Hot\&Co Claims & 0.054$^{**}$ \\ 
  & (0.025) \\ 
 Fact-Checks of Hot\&Co Claims & 0.058 \\ 
  & (0.050) \\ 
 Average Believability of Hot\&Co Claims & 0.060$^{**}$ \\ 
  & (0.024) \\ 
  Exposure $\times$ Fact-Checks - Hot\&Co Claims & $-$0.007 \\ 
  & (0.023) \\ 
  \hline
 Exposure to 5G Claim & $-$0.131$^{***}$ \\ 
  & (0.040) \\ 
 Fact-Check of 5G Claim & 0.014 \\ 
  & (0.148) \\ 
 Average Believability of 5G Claim & $-$0.010 \\ 
  & (0.017) \\ 
 Exposure $\times$ Fact-Checks - 5G Claim & 0.038 \\ 
  & (0.154) \\ 
\hline \\[-1.8ex] 
Observations & 18,314 \\ 
$R^2$ & 0.178 \\
\bottomrule
\end{tabular} 
\end{table} 

\begin{table}[h!] \centering 
  \caption{\textbf{Model 3 mixed-regression results. Each respondents' vaccine decision is predicted based on exposure to false claims and their fact-checks, believability, perceived threat, and vaccination history. Claims are grouped into their respective topic. We include the respondent's residence country as a random effect. Standard errors are presented between parenthesis. Significance marked as $^{*}$ P$<$0.1; $^{**}$ P$<$0.05; $^{***}$ P$<$0.01.}}
  \footnotesize
\begin{tabular}{@{\extracolsep{5pt}}lc} 
\\[-1.8ex]\hline 
\hline \\[-1.8ex] 
 & \multicolumn{1}{c}{\textit{Dependent Variable:}} \\ 
\cline{2-2} 
\\[-1.8ex] & Vaccine Acceptance \\ 
\hline \\[-1.8ex] 
 Constant & $-$1.777$^{***}$ \\ 
  & (0.134) \\ 
 Sex & $-$0.320$^{***}$ \\ 
  & (0.037) \\ 
 Age & $-$0.042$^{***}$ \\ 
  & (0.011) \\ 
 Education & 0.001 \\ 
  & (0.016) \\ 
 Financial Status & 0.118$^{***}$ \\ 
  & (0.020) \\ 
 Health Status & $-$0.062$^{***}$ \\ 
  & (0.021) \\ 
 Past Vaccination & $-$0.110$^{**}$ \\ 
  & (0.053) \\ 
 Past Non-Mandatory Vaccination & 0.625$^{***}$ \\ 
  & (0.078) \\ 
 Past Vaccination $\times$ Past Non-Mandatory Vaccination & 0.213$^{**}$ \\ 
  & (0.087) \\ 
 Perceived Threat & 0.626$^{***}$ \\ 
  & (0.033) \\ 
  \hline
 Exposure to Vaccination-Related Claims & $-$0.164$^{***}$ \\ 
  & (0.020) \\ 
 Fact-Checks of Vaccination-Related Claims & 0.175$^{**}$ \\ 
  & (0.074) \\ 
 Average Believability of  & $-$0.616$^{***}$ \\ 
  & (0.021) \\ 
  Exposure $\times$ Fact-Checks - Vaccination-Related Claims & $-$0.011 \\ 
  & (0.028) \\ 
  \hline
 Exposure to DIY Claims & 0.065$^{***}$ \\ 
  & (0.022) \\ 
 Fact-Checks of DIY Claims & 0.015 \\ 
  & (0.046) \\ 
 Average Believability of DIY Claims & $-$0.025 \\ 
  & (0.028) \\ 
 Exposure $\times$ Fact-Checks - DIY Claims & 0.012 \\ 
  & (0.015) \\ 
  \hline
 Exposure to Hot\&Co Claims & 0.084$^{***}$ \\ 
  & (0.026) \\ 
 Fact-Checks of Hot\&Co Claims & 0.027 \\ 
  & (0.051) \\ 
 Average Believability of Hot\&Co Claims & 0.070$^{***}$ \\ 
  & (0.024) \\ 
 Exposure $\times$ Fact-Checks - Hot\&Co Claims & $-$0.005 \\ 
  & (0.023) \\ 
  \hline
 Exposure to 5G Claim & $-$0.090$^{**}$ \\ 
  & (0.042) \\ 
 Fact-Checks of 5G Claim & 0.036 \\ 
  & (0.152) \\ 
 Average Believability of 5G Claim & $-$0.043$^{**}$ \\ 
  & (0.018) \\ 
 Exposure $\times$ Fact-Checks - 5G Claim & 0.061 \\ 
  & (0.157) \\ 
  \hline \\[-1.8ex] 
Country-Level Random Effects & Yes \\
\hline \\[-1.8ex] 
Observations & 18,314 \\ 
$R^2$ & 0.208\\
\bottomrule
\end{tabular} 
\end{table}

\begin{table}[h!] \centering 
\caption{\textbf{Model 3 elastic regression results. Each respondents' vaccine decision is predicted based on exposure to false claims and their fact-checks, believability, perceived threat, and vaccination history. Claims are grouped into their respective topic.} }
\small
\begin{tabular}{@{\extracolsep{5pt}}lc} 
\toprule
 & \multicolumn{1}{c}{\textit{Dependent Variable:}} \\ 
\cline{2-2} 
\\[-1.8ex] & Vaccine Acceptance \\ 
\hline \\[-1.8ex] 
 Constant & $-$2.210 \\ 
 Sex - Male & 0.248 \\ 
 Sex - Female & $-$0.045 \\ 
 Age & $-$0.050 \\ 
 Education & 0.002 \\ 
 Financial Status & 0.097 \\ 
 Health Status & --- \\ 
 Past Vaccination & $-$0.025 \\ 
 Past Non-Mandatory Vaccination & 0.730 \\ 
  Past Vaccination $\times$ Past Non-Mandatory Vaccination & 0.094 \\ 
 Perceived Threat & 0.709 \\ 
 
  \hline
  
Exposure to Vaccination-Related Claims & $-$0.161 \\
Fact-Checks of Vaccination-Related Claims & 0.133 \\
Average Believability of  & $-$0.582 \\ 
Exposure $\times$ Fact-Checks - Vaccination-Related Claims & --- \\ 

\hline

Exposure to DIY Claims & 0.103 \\ 
Fact-Checks of DIY Claims & 0.002 \\ 
Average Believability of DIY Claims & --- \\ 
Exposure $\times$ Fact-Checks - DIY Claims & 0.012 \\ 

\hline

Exposure to Hot\&Co Claims & 0.046 \\ 
Fact-Checks of Hot\&Co Claims & 0.043 \\ 
Average Believability of Hot\&Co Claims & 0.052 \\ 
Exposure $\times$ Fact-Checks - Hot\&Co Claims & --- \\

\hline

Exposure to 5G Claim & $-$0.116 \\ 
Fact-Checks of 5G Claim & 0.012 \\ 
Average Believability of 5G Claim & $-$0.008 \\ 
Exposure $\times$ Fact-Checks - 5G Claim & 0.024 \\ 

\hline \\[-1.8ex] 
Observations & 18,314 \\ 
\bottomrule
\end{tabular} 
\end{table}


\begin{table}[h!] \centering 
\caption{\textbf{Model 3 lasso regression results. Each respondents' vaccine decision is predicted based on exposure to false claims and their fact-checks, believability, perceived threat, and vaccination history. Claims are grouped into their respective topic.} }
\small
\begin{tabular}{@{\extracolsep{5pt}}lc} 
\toprule
 & \multicolumn{1}{c}{\textit{Dependent Variable:}} \\ 
\cline{2-2} 
\\[-1.8ex] & Vaccine Acceptance \\ 
\hline \\[-1.8ex] 
 Constant & $-$2.267 \\ 
 Sex - Male & 0.294 \\ 
 Sex - Female & 0.000 \\ 
 Age & $-$0.051 \\ 
 Education & 0.002 \\ 
 Financial Status & 0.098 \\ 
 Health Status & --- \\ 
 Past Vaccination & $-$0.024 \\ 
 Past Non-Mandatory Vaccination & 0.739 \\ 
  Past Vaccination $\times$ Past Non-Mandatory Vaccination & 0.087 \\ 
 Perceived Threat & 0.713 \\ 
 
  \hline
  
Exposure to Vaccination-Related Claims & $-$0.162 \\
Fact-Checks of Vaccination-Related Claims & 0.134 \\
Average Believability of  & $-$0.586 \\ 
Exposure $\times$ Fact-Checks - Vaccination-Related Claims & --- \\ 

\hline

Exposure to DIY Claims & 0.103 \\ 
Fact-Checks of DIY Claims & 0.000 \\ 
Average Believability of DIY Claims & --- \\ 
Exposure $\times$ Fact-Checks - DIY Claims & 0.013 \\ 

\hline

Exposure to Hot\&Co Claims & 0.046 \\ 
Fact-Checks of Hot\&Co Claims & 0.043 \\ 
Average Believability of Hot\&Co Claims & 0.054 \\ 
Exposure $\times$ Fact-Checks - Hot\&Co Claims & --- \\

\hline

Exposure to 5G Claim & $-$0.117 \\ 
Fact-Checks of 5G Claim & 0.030 \\ 
Average Believability of 5G Claim & $-$0.007 \\ 
Exposure $\times$ Fact-Checks - 5G Claim & 0.007 \\ 

\hline \\[-1.8ex] 
Observations & 18,314 \\ 
\bottomrule
\end{tabular} 
\end{table}

\begin{figure*}[h!]
\begin{adjustwidth}{-2.25in}{0in}

\caption{\textbf{Country-level exposure to misinformation ({pink}) and fact-checks ({purple}) (S1 Fig.)}}
\begin{flushleft}
The plot indicates the percentage of participants who have seen each claim and its corresponding fact-check in a specific country covered by our study. The numbers are calculated after post-stratification weighting by the process of raking. The radial axis represents the percentages ranging from 0 to 100. In the angular axis, distinct claims representing similar notions, i.e., vaccination-related claims, are arranged together. The titles of the plots are each country’s ISO 3166-1 alpha-3 codes.
\end{flushleft}
    \centering
    \includegraphics[width=.73\linewidth]{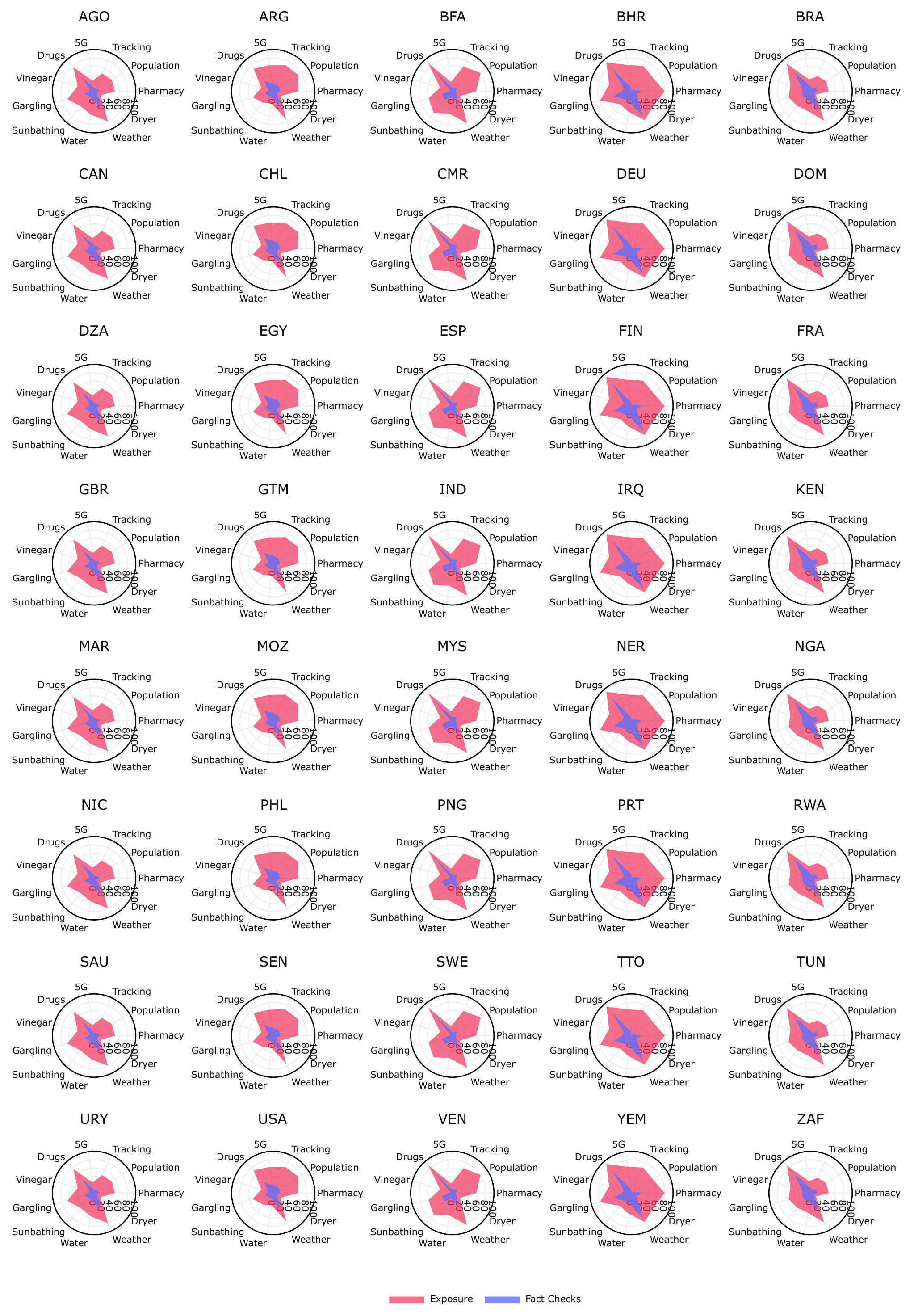}
    \label{fig:allcountries}
\end{adjustwidth}
\end{figure*}

%% file: main.bbl
\begin{thebibliography}{10}

\bibitem{kwon2017plos}
Kwon S, Cha M, Kyomin J.
\newblock Rumor detection over varying time windows.
\newblock PloS one. 2017;12(1):e0168344.

\bibitem{shao2018spread}
Shao C, Ciampaglia GL, Varol O, Yang KC, Flammini A, Menczer F.
\newblock The spread of low-credibility content by social bots.
\newblock Nature communications. 2018;9(1):1--9.

\bibitem{science18vosoughi}
Vosoughi S, Roy D, Aral S.
\newblock The spread of true and false news online.
\newblock science. 2018;359(6380):1146--1151.

\bibitem{flickr2009www}
Cha M, Mislove A, Gummadi KP.
\newblock {A Measurement-driven Analysis of Information Propagation in the
  Flickr Social Network}.
\newblock In: Proceedings of the International World Wide Web Conference; 2009.

\bibitem{zarocostas2020fight}
Zarocostas J.
\newblock How to fight an infodemic.
\newblock The Lancet. 2020;395(10225):676.

\bibitem{who20infodemic}
WHO, WHO Novel Coronavirus (2019-nCoV) Situation Report 13; 2020.

\bibitem{kh20saltwater}
The Korea Herald, River of Grace Community Church sprays saltwater into
  worshippers' mouths; 2020.

\bibitem{nyt20burning}
New York Times, Burning Cell Towers, Out of Baseless Fear They Spread the
  Virus; 2020.

\bibitem{vinck2019institutional}
Vinck P, Pham PN, Bindu KK, Bedford J, Nilles EJ.
\newblock Institutional trust and misinformation in the response to the
  2018--19 Ebola outbreak in North Kivu, DR Congo: a population-based survey.
\newblock The Lancet Infectious Diseases. 2019;19(5):529--536.

\bibitem{whohpv}
WHO. Global Advisory Committee on Vaccine Safety, 5-6 June 2019; 2020.

\bibitem{benecke2019anti}
Benecke O, DeYoung SE.
\newblock Anti-vaccine decision-making and measles resurgence in the United
  States.
\newblock Global Pediatric Health. 2019;6:2333794X19862949.

\bibitem{orenstein2017simply}
Orenstein WA, Ahmed R. Simply put: vaccination saves lives; 2017.

\bibitem{draulans2020finally}
Draulans D.
\newblock Finally, a virus got me. Scientist who fought Ebola and HIV reflects
  on facing death from COVID-19.
\newblock Sci Mag [May 8, 2020]. 2020;.

\bibitem{kwon2013icdm}
Kwon S, Cha M, Jung K, Chen W, Wang Y.
\newblock {Prominent Features of Rumor Propagation in Online Social Media}.
\newblock In: Proceedings of the IEEE International Conference on Data Mining;
  2013.

\bibitem{ma2016ijcai}
Ma J, Gao W, Mitra P, Kwon S, Jansen BJ, Wong KF, et~al.
\newblock {Detecting rumors from microblogs with recurrent neural networks}.
\newblock In: Proceedings of the International Joint Conference on Artificial
  Intelligence; 2016.

\bibitem{ma2018rumor}
Ma J, Gao W, Wong KF.
\newblock Rumor detection on twitter with tree-structured recursive neural
  networks.
\newblock Association for Computational Linguistics; 2018.

\bibitem{nyt20twitter}
The New York Times, Twitter Will Turn Off Some Features to Fight Election
  Misinformation; 2020.

\bibitem{verge20qanon}
The Verge, Facebook completely bans QAnon and labels it a `militarized social
  movement'; 2020.

\bibitem{fbr}
IBS DSG. Facts Before Rumors; 2020.

\bibitem{info20global}
Islam MS, Sarkar T, Khan SH, Kamal AHM, Hasan SMM, Kabir A, et~al.
\newblock COVID-19?Related Infodemic and Its Impact on Public Health: A Global
  Social Media Analysis.
\newblock The American Journal of Tropical Medicine and Hygiene. 07 Oct
  2020;103(4):1621 -- 1629.

\bibitem{twit20explora}
Shahi GK, Dirkson A, Majchrzak TA.
\newblock An exploratory study of COVID-19 misinformation on Twitter.
\newblock Online Social Networks and Media. 2021;22:100104.

\bibitem{cinelli2020covid}
Cinelli M, Quattrociocchi W, Galeazzi A, Valensise CM, Brugnoli E, Schmidt AL,
  et~al.
\newblock The covid-19 social media infodemic.
\newblock Scientific Reports. 2020;10(1):1--10.

\bibitem{kouzy2020coronavirus}
Kouzy R, Abi~Jaoude J, Kraitem A, El~Alam MB, Karam B, Adib E, et~al.
\newblock Coronavirus goes viral: quantifying the COVID-19 misinformation
  epidemic on Twitter.
\newblock Cureus. 2020;12(3).

\bibitem{whomythbusters}
WHO. Coronavirus disease (COVID-19) advice for the public: Mythbusters; 2020.

\bibitem{covidpoynter}
Poynter. COVID-19: Poynter Resources; 2020.

\bibitem{palan2018prolific}
Palan S, Schitter C.
\newblock Prolific. ac—A subject pool for online experiments.
\newblock Journal of Behavioral and Experimental Finance. 2018;17:22--27.

\bibitem{kachanoff2020realistic}
Kachanoff FJ, Bigman YE, Kapsaskis K, Gray K.
\newblock Measuring Realistic and Symbolic Threats of COVID-19 and Their Unique
  Impacts on Well-Being and Adherence to Public Health Behaviors.
\newblock Social Psychological and Personality Science.
  0;0(0):1948550620931634.
\newblock doi:{10.1177/1948550620931634}.

\bibitem{facebookreport}
Facebook. Facebook Reports First Quarter 2020 Results; 2020.

\bibitem{schneider2019s}
Schneider D, Harknett K.
\newblock What’s to like? Facebook as a tool for survey data collection.
\newblock Sociological Methods \& Research. 2019; p. 0049124119882477.

\bibitem{pham2019online}
Pham KH, Rampazzo F, Rosenzweig LR.
\newblock Online surveys and digital demography in the developing world:
  Facebook users in Kenya.
\newblock arXiv preprint arXiv:191003448. 2019;.

\bibitem{ribeiro2020biased}
Ribeiro FN, Benevenuto F, Zagheni E.
\newblock How Biased is the Population of Facebook Users? Comparing the
  Demographics of Facebook Users with Census Data to Generate Correction
  Factors.
\newblock arXiv preprint arXiv:200508065. 2020;.

\bibitem{kalton2003weighting}
Kalton G, Flores-Cervantes I.
\newblock Weighting methods.
\newblock Journal of official statistics. 2003;19(2):81.

\bibitem{fb2020biased}
Ribeiro FN, Benevenuto F, Zagheni E.
\newblock How Biased is the Population of Facebook Users? Comparing the
  Demographics of Facebook Users with Census Data to Generate Correction
  Factors.
\newblock In: Proceedings of12th ACM Conference on Web Science. WebSci '20;
  2020. p. 325–334.

\bibitem{paolacci2010running}
Paolacci G, Chandler J, Ipeirotis PG.
\newblock Running experiments on amazon mechanical turk.
\newblock Judgment and Decision making. 2010;5(5):411--419.

\bibitem{ross2010crowdworkers}
Ross J, Irani L, Silberman MS, Zaldivar A, Tomlinson B.
\newblock Who are the crowdworkers? Shifting demographics in Mechanical Turk.
\newblock In: CHI'10 extended abstracts on Human factors in computing systems;
  2010. p. 2863--2872.

\bibitem{trump2020weather}
CNN, Prestigious scientific panel tells White House coronavirus won't go away
  with warmer weather; 2020.

\bibitem{horby2020effect}
Horby P, Mafham M, Linsell L, Bell JL, Staplin N, Emberson JR, et~al.
\newblock Effect of Hydroxychloroquine in Hospitalized Patients with Covid-19.
\newblock New England Journal of Medicine. 2020;.

\bibitem{bozo2020hydro}
The New York Times, Bolsonaro Hails Anti-Malaria Pill Even as He Fights
  Coronavirus; 2020.

\bibitem{trump2020hydro}
Vox, Trump’s reckless promotion of hydroxychloroquine to fight coronavirus,
  explained; 2020.

\bibitem{xaudiera2020ibuprofen}
Xaudiera S, Cardenal AS.
\newblock Ibuprofen narratives in five european countries during the COVID-19
  Pandemic.
\newblock Harvard Kennedy School Misinformation Review. 2020;1(3).

\bibitem{eu2015hliteracy}
Sørensen K, Pelikan JM, Röthlin F, Ganahl K, Slonska Zea.
\newblock Health literacy in Europe: comparative results of the European health
  literacy survey (HLS-EU).
\newblock European Journal of Public Health. 2015;25(6):1053--1058.

\bibitem{lorini2018}
Lorini C, Ierardi F, Bachini L, Donzellini M, Gemmi F, Bonaccorsi G.
\newblock The Antecedents and Consequences of Health Literacy in an Ecological
  Perspective: Results from an Experimental Analysis.
\newblock Int J Environ Res Public Health. 2018;15(4):798.

\bibitem{corscadden2018factors}
Corscadden L, Levesque J, Lewis V, Strumpf E, Breton M, Russell G.
\newblock Factors associated with multiple barriers to access to primary care:
  an international analysis.
\newblock International journal for equity in health. 2018;17(1):1--10.

\bibitem{meyer2013inequities}
Meyer SB, Luong TC, Mamerow L, Ward PR.
\newblock Inequities in access to healthcare: analysis of national survey data
  across six Asia-Pacific countries.
\newblock BMC health services research. 2013;13(1):238.

\bibitem{zar2020challenges}
Zar HJ, Dawa J, Fischer GB, Castro-Rodriguez JA.
\newblock Challenges of COVID-19 in children in low-and middle-income
  countries.
\newblock Paediatric Respiratory Reviews. 2020;.

\bibitem{med2015dist}
Carr-Hill R, Currie E.
\newblock What explains the distribution of doctors and nurses in different
  countries, and does it matter for health outcomes?
\newblock J Adv Nurs. 2013;69(11):2525--2537.

\bibitem{hudspeth2017health}
Hudspeth J, Morse M.
\newblock Health Information and Global Health Inequity: Point-of-Care
  Knowledge Systems as a Foundation for Progress.
\newblock Journal of general internal medicine. 2017;32(5):572--575.

\bibitem{poushter2018social}
Poushter J, Bishop C, Chwe H.
\newblock Social media use continues to rise in developing countries but
  plateaus across developed ones.
\newblock Pew Research Center. 2018;22:2--19.

\bibitem{puri2020social}
Puri N, Coomes EA, Haghbayan H, Gunaratne K.
\newblock Social media and vaccine hesitancy: new updates for the era of
  COVID-19 and globalized infectious diseases.
\newblock Human Vaccines \& Immunotherapeutics. 2020; p. 1--8.

\bibitem{johnson2020online}
Johnson NF, Vel{\'a}squez N, Restrepo NJ, Leahy R, Gabriel N, El~Oud S, et~al.
\newblock The online competition between pro-and anti-vaccination views.
\newblock Nature. 2020; p. 1--4.

\bibitem{vergeytvacc}
The Verge, YouTube will remove videos with COVID-19 vaccine misinformation;
  2020.

\bibitem{vergefbvacc}
The Verge, Facebook announces ban on anti-vaccination ads; 2020.

\bibitem{5g20guardian}
'Quite frankly terrifying': How the QAnon conspiracy theory is taking root in
  the UK; 2020.

\bibitem{ireland1968contingency}
Ireland CT, Kullback S.
\newblock Contingency tables with given marginals.
\newblock Biometrika. 1968;55(1):179--188.

\bibitem{Battaglia2009Practical}
Battaglia MP, Hoaglin DC, Frankel MR.
\newblock Practical Considerations in Raking Survey Data.
\newblock Survey Practice. 2009;2(5).
\newblock doi:{10.29115/SP-2009-0019}.

\bibitem{ncd19book}
Grosskurth H.
\newblock The Burden of Non-communicable Diseases in Low- and Middle-Income
  Countries.
\newblock In: Atkinson K, Mabey D, editors. Revolutionizing Tropical Medicine.
  John Wiley \& Sons; 2019.

\bibitem{fisher1993social}
Fisher RJ.
\newblock Social desirability bias and the validity of indirect questioning.
\newblock Journal of consumer research. 1993;20(2):303--315.

\bibitem{larsen2020survey}
Larsen M, Nyrup J, Petersen MB.
\newblock Do survey estimates of the public’s compliance with COVID-19
  regulations suffer from social desirability bias?
\newblock Journal of Behavioral Public Administration. 2020;3(2).

\bibitem{kalimeri2019evaluation}
Kalimeri K, Beir{\'o} MG, Bonanomi A, Rosina A, Cattuto C.
\newblock Evaluation of biases in self-reported demographic and psychometric
  information: traditional versus Facebook-based surveys.
\newblock arXiv preprint arXiv:190107876. 2019;.

\end{thebibliography}
